# The (human) respiratory rate at rest

Wolfgang Schramm   *wolfgang.schramm@meduniwien.ac.at*

Austria 1180 Vienna, Schulgasse 62/11   https://www.schramm.cc

## Summary

All schoolchildren know how often they breathe, but even experts don't know exactly why. The aim of this publication is to develop a model of the resting spontaneous breathing rate using physiological, physical and mathematical methods with the aid of the principle that evolution pushes physiology in a direction that is as economical as possible. The respiratory rate then follows from an equation with the parameters $CO_2$-production rate of the organism, resistance, static compliance and dead space of the lungs, the inspiration duration: expiration duration - ratio and the end-expiratory $CO_2$ fraction. The derivation requires exclusively secondary school mathematics. Using the example of an adult human or a newborn child, data from the literature then result in normal values for their breathing rate at rest. The reason for the higher respiratory rate of a newborn human compared to an adult is the relatively high $CO_2$-production rate together with the comparatively low compliance of the lungs. A side result is the fact that the common alveolar pressure throughout the lungs and the common time constant is a consequence of the economical principle as well. Since the above parameters are not human-specific, there is no reason to assume that the above equation could not also be applicable to many animals breathing through lungs within a thorax, especially mammals. Not only physiology and biology, but also medicine, could benefit: Applicability is being discussed in pulmonary function diagnostics, including pathophysiology. However, the present publication only claims to be a theoretical concept for the spontaneous quiet breathing rate. In the absence of comparable animal data, this publication is intended to encourage further scientific tests.

**Keywords:** respiratory rate, evolutionary economy, respiratory physiology, volume flow pattern, $CO_2$-production rate, static compliance, airway resistance, end-expiratory $CO_2$ fraction, time constant; alveolar pressure

**Mathematics Subject Classification:** 92C30, 92C05

**Funding:** This project was not funded.

## Introduction

The aim of this publication is to develop a model of the resting spontaneous breathing rate that is as simple as possible using physiological, physical and mathematical methods with the aid of the principle that evolution pushes physiology in a direction that is as economical as possible. The respiratory rate $F_{opt}$ then follows thereof from an equation with the parameters $CO_2$ production rate $\dot{V}CO_2$ of the organism, resistance $R$, static compliance $C$ and dead space $V_D$ of the lungs, the inspiration duration: expiration duration ratio $I:E$ and the end-expiratory $CO_2$-fraction $F_{Et}CO_2$. In order to make this publication readable for many, only elementary analysis without numerical methods shall be used for this main result which is also helpful for intuition.

In the following, well-known terms and facts of respiratory physiology are sometimes repeated with full intent in order to transcribe them into a physical language with mathematical aids. This is necessary because in respiratory physiology, for example, an inflection point has a different meaning than in mathematics, and if compliance is defined as a difference quotient as usual, this is introduced below as a differential quotient. Even physical units can sometimes not be translated in medical publications, or only with difficulty (Schramm 2010), for example the unit *gm* for a work in studies (Otis 1950; Crosfill 1961) which use clinical data to determine the breathing rate at rest. A certain analogy between lung physiology and electrical engineering (Campbell 1963) can be helpful here (table 1).

| **Pulmonary physiology** | **Electrical engineering** |
|---|---|
| Volume $V$ | Charge $Q$ |
| Pressure difference $\Delta p$ | Voltage difference $\Delta U$ |
| Volume flow $dV/dt$ | Current $I = dQ/dt$ |
| Airway resistance $R = \Delta p/(dV/dt)$ | Resistance $R = \Delta U/I$ (=Ohm's law) |
| Resistance pressure $\Delta p = R \cdot (dV/dt)$ | Voltage across a resistor $\Delta U = R \cdot (dQ/dt)$ |
| Static compliance $C(p) = dV/dp$ | Capacity $C = \Delta Q/\Delta U$ |
| Compliance pressure $\Delta p = \Delta V/C$ | Voltage across a capacitor $\Delta U = \Delta Q/C$ |
| Time constant $\tau = R \cdot C$ | Time constant $\tau = R \cdot C$ |
| Compliance work $W_C = V_T^2/(2C)$ | Energy stored in a capacitor $W = Q^2/(2C)$ |
| Passive expiration function at rest $V(t) = V_T \cdot e^{-t/RC}$ | RC-discharging function $Q(t) = Q_0 \cdot e^{-t/RC}$ |
| Passive expiration pressure difference at rest $\Delta p(t) = \Delta p_0 \cdot e^{-t/RC}$ | RC-discharging Voltage difference at $R$: $\Delta U(t) = \Delta U_0 \cdot e^{-t/RC}$ |
| Serial resistances $R = R_1 + R_2$ | Serial resistances $R = R_1 + R_2$ |
| Parallel resistances $1/R = 1/R_1 + 1/R_2$ | Parallel resistances $1/R = 1/R_1 + 1/R_2$ |
| Resistance power $P_R = (dV/dt)^2 \cdot R$ | Resistance power $P = I^2 \cdot R$ |
| Parallel compliances $C = C_1 + C_2$ | Parallel capacitances $C = C_1 + C_2$ |

**Table 1:** Analogy ("in a sense even an isomorphism") between pulmonary physiology and electrical engineering. Any line on the left that contains the airway resistance $R$, requires laminar flow. In the case of a capacitor, the change in charge $\Delta Q$ leads to a proportional change in voltage $\Delta U = \Delta Q/C$, since the capacitance $C$ usually does not change during loading. However, the compliance of the lungs and thorax is not constant but a function $C(p)$ of the pressure $p$ which is why, in contrast to electrical engineering, the transition from the differential quotient $\Delta V/\Delta p$ to the derivative $dV/dp$ is relevant in lung physiology. However, under spontaneous breathing at rest $C(p)$ may by considered as constant.
2

# Total work of breathing

A survival advantage, especially in the energy balance, is certainly not a disadvantage in times of low food resources and consequently an advantage over living beings that are less economical in this regard. Therefore, the following "postulate of evolutionary economy" is assumed:

**Postulate:** Evolution pushes physiology in a direction that is as economical as possible. (P)

Consequently, it can be assumed that evolution has minimized the total respiratory power $P_{resp}$ (work/time) as well. Hence, many authors (Otis 1950; Kerem 1996; Noël 2019) have assumed that the breathing power must be as small as possible during spontaneous breathing. In purely physical terms, the work of breathing is composed of
1. Compliance work $W_C$ (= work to expand the lungs against the lung and chest wall elastic forces)
2. Resistance work (= work to overcome the airway resistance $R$)
3. Tissue resistance work ( = work to overcome the viscosity of the lung and chest wall structures)
4. Acceleration work of the lungs, chest wall and breathing gas

additively.

The power required for acceleration is, however, only relevant in diving medicine during deep dives (Dwyer 1977) together with large respiratory minute volumes $V_T \cdot F$ (= tidal volume times respiratory rate), but negligible at rest under one atmosphere of air pressure (Rohrer 1925), as a simple calculation at the end will show as well. The tissue resistance work will be discussed below together with the airway resistance $R$.

The ambient pressure $p_B$ is usually chosen as the reference pressure $p_B = 0$ in (respiratory) physiology and biology. Consequently you can replace $p_B$ with zero nearly everywhere in this paper (except e.g. in the ideal gas equation or the definition of the gas fraction).

**Definition:** The pleural pressure $p_P(t)$ is the pressure within the narrow space (=pleural space) between lungs- and chest wall pleura.

**Definition:** The pressure within all the alveoli (= alveolar space) is denoted alveolar pressure $p_A(t)$.

The next but one subsection shows that $p_A(t)$ is a common pressure for all the alveoli which lastly follows from (P) as well.

There is no doubt among physiologists that spontaneous expiration at rest is passive (Levitzky 2018; Lumb 2016a). Since this is also the case under controlled positive pressure ventilation and, moreover, volumes, pressures and volume flows can be measured very easily under mechanical ventilation, not only analogies between pulmonary physiology and electrical engineering, but also analogies between spontaneous breathing and mechanical ventilation will often be helpful in the following.

During positive pressure mechanical ventilation the ambient pressure in front of the lungs (mouth, nose or trachea) seemingly changes which is responsible for the volume flow $\frac{dV}{dt}$.

During spontaneous breathing however, it is the changing pleural pressure $p_P$ which is responsible for the volume flow $\frac{dV}{dt}$ and the alveolar pressure $p_A(t)$ changes the sign with respect to the ambient pressure $p_B$ between inspiration and expiration which is not the case under positive pressure ventilation. This is ultimately why this type of mechanical ventilation is called positive pressure ventilation. Side-note: During ventilation with the historical iron lung a (negative) ambient pressure of the eintire body except the head is responsible for the volume flow which actually is more physiologic.

**Definition:** During positive pressure ventilation the pressure difference between the (seemingly changing ambient) pressure in front of the lungs (e.g. within the ventilation device: endotracheal tube, laryngeal mask, …) respectively face (e.g. face mask, …) and the alveolar pressure $p_A$ is called (positive) ventilation pressure $p_V(t)$.

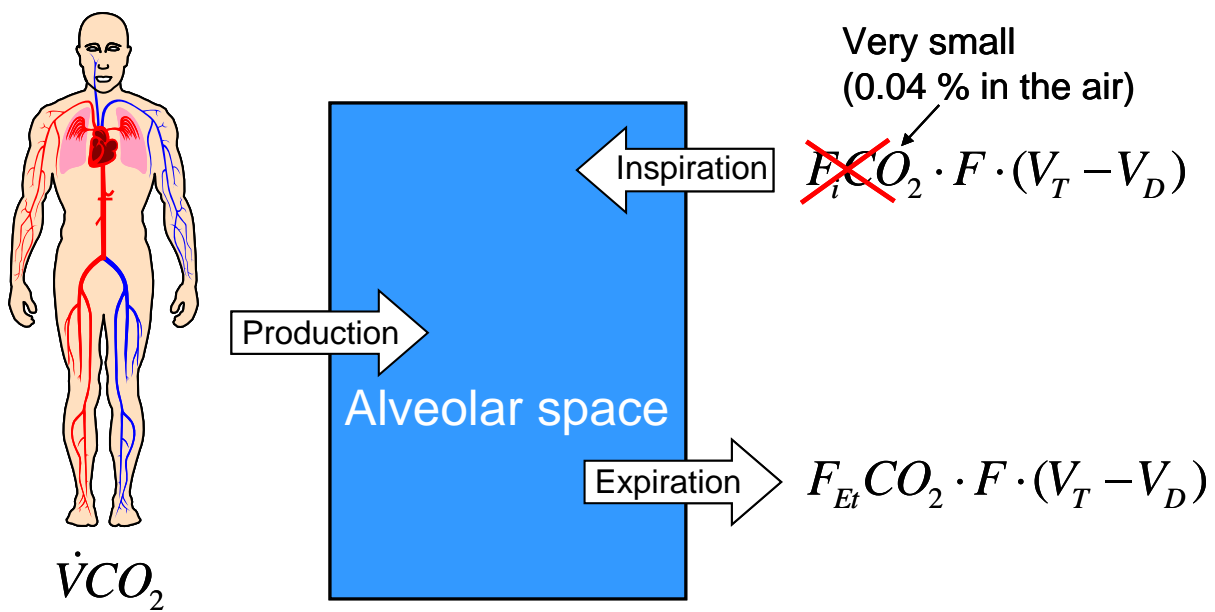

**Figure 1**: $CO_2$ balance if the (human) organism is viewed as a black box. With $F_iCO_2$ respectively $F_{Et}CO_2$ the inspiratory or end-tidal $CO_2$ fraction is referred to and $F \cdot (V_T - V_D)$ is the alveolar ventilation. The $CO_2$-production rate is denoted $\dot{V}CO_2$.

The $CO_2$-partial pressure within the organism is responsible for the respiratory drive at least under one atmosphere of air pressure. Therefore, the $CO_2$ balance is being discussed first which will lead to a side condition for the calculation of the respiratory rate needed later.

### *The $CO_2$ balance of the organism and the alveolar ventilation*

A slight change in the $O_2$-partial pressure of the breathing air at sea level, e.g. due to weather conditions, undoubtedly has a negligible influence on the tidal volume $V_T$ and the breathing rate $F$. However, this changes with increasing altitude (Duffin 2007; Weil 1970). On Mt. Everest the respiratory minute volume $F \cdot V_T$ even takes on considerable size (Pugh 1957; Weil 1970) which is why an air pressure $p_B$ of 760 mmHg is assumed in the following. If the organism is viewed as a black box (figure 1), the $CO_2$ balance can be summarized in the equation:

$$CO_2 \text{ production rate} + CO_2 \text{ inspiration} - CO_2 \text{ expiration} = 0$$

if the $CO_2$ excretion via the kidneys (bicarbonate or in physical solution) is neglected, especially since the latter is at least a factor of $10^3$ less than the $CO_2$ output via expiration.

**Definition:** The anatomical dead space, such as the nose, mouth, trachea and the bronchial tree, is the part of the tidal volume $V_T$ that does not participate in gas exchange which is predetermined by the anatomy. If, in pathophysiology, there are additional components that do not participate in gas exchange, the entire dead space $V_D$ is then referred to as total dead space.

The part of the minute volume $F \cdot V_T$ that takes part in gas exchange within the lungs is accordingly $F \cdot (V_T - V_D)$ and is referred to as alveolar ventilation.

The fraction $F_{Gas} := \frac{p_{Gas}}{p_B}$ of a gas is defined as the quotient of the partial pressure $p_{Gas}$ of any gas and the total pressure $p_B$ which is always $760 mmHg$ here. Let $F_i CO_2$ be the inspiratory, $F_{Et} CO_2$ the end-tidal $CO_2$-fraction (figure 2) and $\dot{V}CO_2$ the $CO_2$ production rate (in a volume / time unit), then the above $CO_2$ balance equation can be rewritten as follows:

$$\dot{V}CO_2 + F_i CO_2 \cdot F \cdot (V_T - V_D) - F_{Et} CO_2 \cdot F \cdot (V_T - V_D) = 0$$

The inspiratory $CO_2$-fraction $F_i CO_2$, compared to the physiological expiratory $CO_2$-fraction $F_{Et} CO_2$ (in humans approx. 5.2%, see table 2) is at least a factor of $10^2$ (in humans approx. 130) lower. Therefore, the inspiratory proportion in the above equation can be ignored. Contrary to the mathematical and physical convention, approximations will continue to be noted with an equal sign in the following. Then the following "alveolar ventilation equation" applies (West 2007):

$$\frac{p_{Et} CO_2}{p_B} =: F_{Et} CO_2 = \frac{\dot{V}CO_2}{F \cdot (V_T - V_D)} \qquad (1)$$

Physiologically, the $p_{Et} CO_2$ is assumed to be representative for the alveolar gas. Therefore, the "inverse proportionality" between the alveolar partial pressure of $CO_2$ and the alveolar ventilation $F \cdot (V_T - V_D)$ is often graphically depicted in physiology textbooks, usually without citing Eq. (1), for example on p. 514 figure 40-5 in the book of Hall (Hall 2020). Because of the extremely rapid diffusion of $CO_2$ through the alveolar membrane as a result of the very good solubility of $CO_2$ in aqueous body fluids (Fenn 1946; Weingarten 1990), the arterial $pCO_2$ correlates very good with the alveolar partial pressure of $CO_2$ which is apparently why Chambers et al. (Chambers 2019) on p.46 replaced the index Et (=end-tidal) in equation (1) with the arterial one.

The $CO_2$ production rate $\dot{V}CO_2 = \dot{V}O_2 \cdot RQ$ depends on the nutritional-dependent respiratory quotient $RQ$ and on the $O_2$-consumption rate $\dot{V}O_2$ and the latter in the resting state (due to the aerobic metabolism in this state) on the caloric equivalent and the constant basal metabolic rate. Admittedly the basal metabolic rate includes the resting respiratory power $P_{resp}$ as well and $P_{resp}$ surely depends on $F$. However, in the resting state $P_{resp}$ makes up only a small fraction of the basal metabolic rate. (Using the example of humans, this will be shown later.) Moreover from the postulate (P) $dP_{resp}/dF = 0$ at the resting respiratory rate $F_{opt}$ is to be expected. Both together justify the assumption that $\dot{V}CO_2$ is (almost) constant within a neighbourhood of $F_{opt}$ which therefore will be assumed in this paper.

The dead space $V_D$ given by the anatomy (or pathophysiology) and age can be regarded as a constant in the following context and the physiological $CO_2$ homeostasis within the organism is responsible for a largely constant end-tidal $CO_2$-fraction $F_{Et}CO_2$.

The respiratory rate $F$ in Eq. (1) can vary by adjusting the tidal volume $V_T$ in order to maintain a physiological $CO_2$ homeostasis in the organism. It is known, however that the breathing rate of the organism at rest on average within a range, depending on age, is largely predetermined by biology, physiology or pathophysiology. The question therefore arises which parameters determine the breathing rate $F_{opt}$ at rest and thus also determine the tidal volume $V_{T,opt}$ as well. Interestingly, this question can be answered with relatively simple aids, at least under physiological conditions.

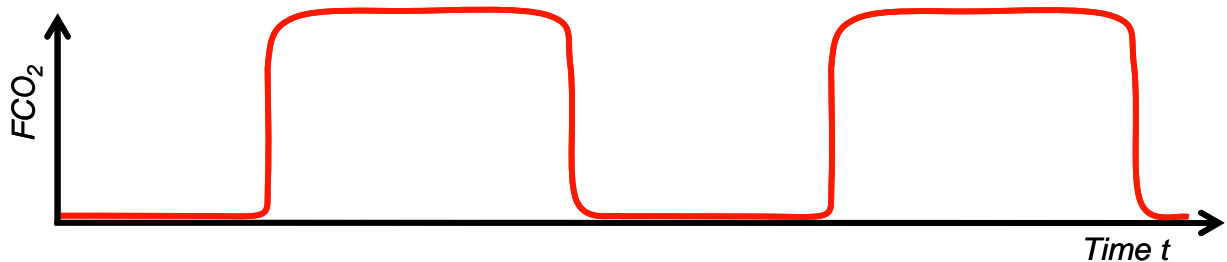

**Figure 2:** Idealized course of an end-tidal $CO_2$ curve (Weingarten 1990) under ventilation or spontaneous breathing. This function $FCO_2(t) = p_{CO_2}(t)/p_B$ is similar to a periodic step function. The upper tangent parallel to the $t$-axis corresponds to the value $F_{Et}CO_2$ and the lower one to the value $F_iCO_2$.

## *Minimum of the respiratory compliance work and power*

The elastic work to expand lungs and chest wall is being discussed first. Let $TLC$ be the total lung capacity, $RV$ the residual volume and $FRC$ the functional residual capacity ($=RV+$ expiratory reserve volume) which is the physiologic resting (or balance) position volume in the lungs at the end of expiration (figure 3).

Quasi-static means that the parameters $X \in \{V, p_A, p_V, ...\}$ are independent of the time $t$: $\frac{dX}{dt} = 0$. Hence, pressure changes which are caused by the volume flow $\frac{dV}{dt}$ should be as small as possible. From Eq. (9), Eq. (16) or Eq. (17) then follows $p_V = p_A$. For this reason the index A respectively V will be suppressed within this subsection. Since the pressure $p$ and the volume $V$ pushed into or released from the lungs can easily be measured during positive pressure mechanical ventilation, the function $V(p)$ of the intrapulmonary volume $V$ as a function of the ventilation pressure $p$ can be easily ascertained quasi-statically for $V(p) \geq FRC$ and $V(p)$ is then referred to as static (figure 3).

**Definition:** A *PEEP* is that pressure which is positive compared to the ambient pressure $p_B$ in the alveolar space at the end of expiration: $PEEP := p_A(t)\big|_{t=\text{end of expiration}} - p_B$.

A *PEEP* is usually specified on a ventilator using an adjusting screw or can be caused pathophysiologically by an increased airway resistance in the dead space, for example in chronic obstructive pulmonary disease (COPD) patients or during an asthma attack, but also by resistances in the ventilation device (endotracheal tube, laryngeal mask, ...). As shown below, using the example of the human data from table 2, there is (almost) no *PEEP* under physiological quiet breathing. Consequently (with the above convention) physiologically $FRC = V(p_B)$ applies.

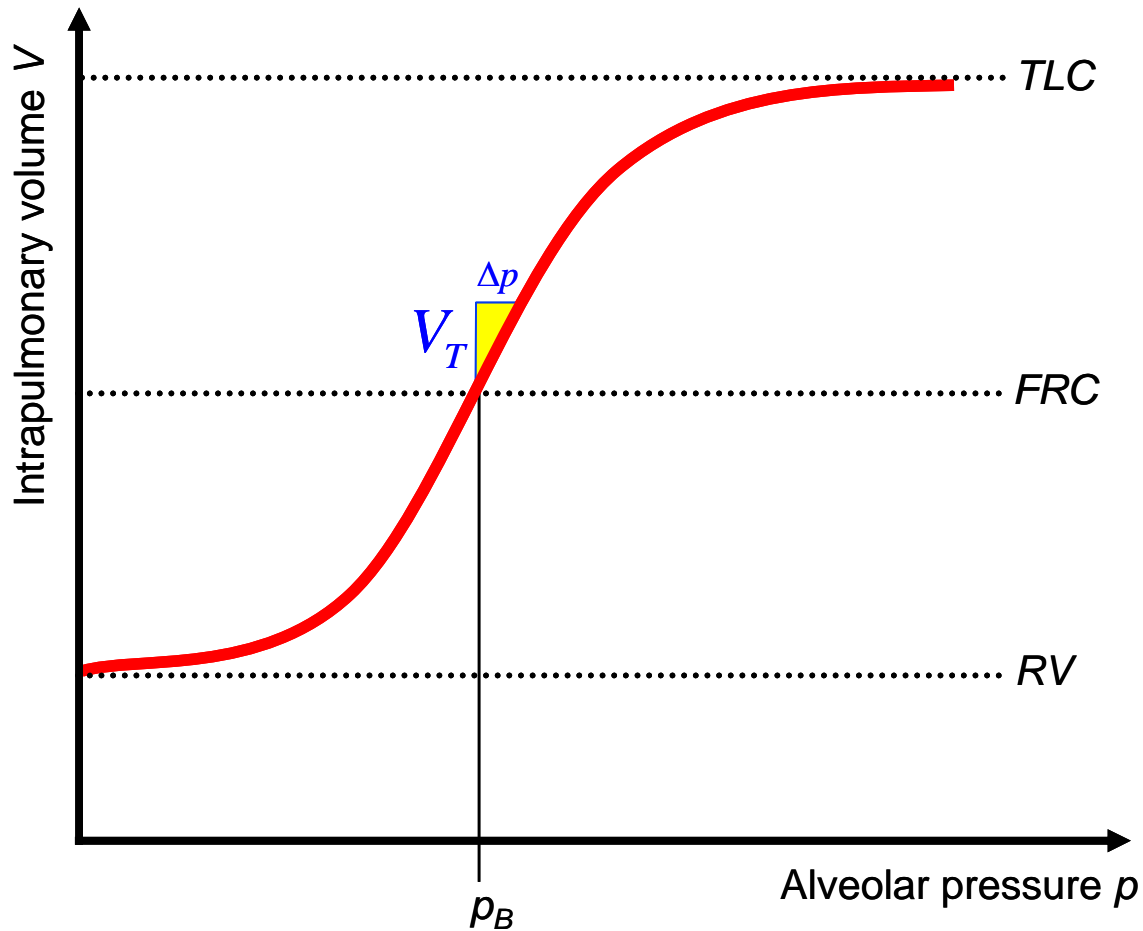

**Figure 3:** *TLC* is the total lung capacity, *RV* the residual volume, $FRC = V(p_B)$ the physiologic functional residual capacity and $p_B$ is the ambient pressure. If under very slow ventilation, i.e. quasi-statically, the tidal volume $V_T$ is pushed into the lungs starting from the *FRC*, then the alveolar (=intrapulmonary) pressure $p$ $(= p_A)$ increases by $\Delta p$ and the intrapulmonary volume follows this strictly increasing function $V(p)$ which is called static. Both elastic structures of the lungs and the thorax surrounding the lungs are responsible for the course of this function.

The function $V(p)$ in figure 3 was therefore created with particularly slow inspiration or expiration. You cannot breathe in beyond the *TLC* and you cannot breathe out below the *RV*. Consequently, an S-shaped course (Rahn 1946; Tepper 2015; Bryan 2011) (s. figure 3) of the function $V(p)$ is to be expected which therefore has exactly one inflection point. Many publications e.g. (Grinnan 2005) as well as textbooks (Kacmarek 2019; Campbell 1963;

Exline 2020), on the other hand, define on $V(p)$ an upper and a lower inflection point which, however, are not inflection points in the mathematical sense and are therefore not referred to as inflection points in the following. The mathematical inflection point of the function $V(p)$ is in any case depicted in textbooks (Chatburn 2003; Bryan 2011) and publications (Rahn 1946) close to the $FRC$.

Both elastic structures of the lungs and the thorax surrounding the lungs are responsible for the course of $V(p)$. Starting from the resting position $FRC$, it is physically evident that the volume of the lungs (strictly) increases while the pressure within the lungs (strictly) increases and the other way around: Starting from the $FRC$, the volume of the lungs (strictly) decreases while the pressure within the lungs (strictly) decreases. Therefore, as can be seen from figure 3, $V(p)$ increases strictly and consequently, the inverse function $p(V)$ exists.

From a physical point of view, changing any pressure $p$ on a null set would have no effect on the volume $V$ and all other parameters (s. table 1) in this paper are in some sense derived from these two and the time $t$. Hence, let all occurring variables be sufficient smooth which is biologically meaningful as well. For example, let us in the following assume that $V(p)$ is at least twice continuously differentiable (i.e $V(p) \in C^2$). In physics, especially thermodynamics this assumption is usually accepted as well.

The location of the inflection point of $V(p)$ is being discussed now. The slope, i.e. the first derivative of $V(p)$, is referred to $C(p) := \dfrac{dV}{dp}$ here as static compliance. (Note: Often, however, in publications or textbooks the difference quotient $\dfrac{\Delta V}{\Delta p}$ is defined as the static compliance.)

If, as an exception, mechanical ventilation is carried out without $PEEP$, then the pressure increases during the inspiration phase, starting from $p = p_B$ with a correspondingly slow ventilation, depending on $p(V)$, in order to passively fall back to $p = p_B$ in the expiration phase. With positive pressure ventilation, the tidal volume $V_T$ is pushed into the lungs during each inspiration. Since $V(p_B) = FRC$, the inspiratory quasi-static work of breathing $W_C$ per breath which is referred to here as compliance work, is:

$$W_C = \int_{FRC}^{FRC+V_T} (p - p_B) dV \qquad (2)$$

because the product of a volume and a pressure difference has the physical unit of a work.
The following few lines will justify the assumption that $C(p)$ is approximately constant and highest within a neighbourhood of the $FRC$. The tangent of $V(p)$ at $V = FRC$ is the straight line equation:

$$V - FRC = (p - p_B) \cdot C \qquad (3)$$

The slope $C$ of this tangent is of course $\dfrac{dV}{dp}(p_B) := \lim_{p \to p_B} \dfrac{V(p) - FRC}{p - p_B}$. With this linearization the inspiratory compliance work (2) can be rewritten:

$$W_C = \int_{FRC}^{FRC+V_T} (\dfrac{V - FRC}{C}) dV = \dfrac{V_T^2}{2C} \qquad (4)$$

which (approximately) corresponds to the yellow colored triangle in figure 3 with the area $\frac{V_T \cdot \Delta p}{2}$ as usually given in textbooks (Hall 2020). (This work is equivalent to the energy content $\frac{Q \cdot \Delta U}{2}$ of a capacitor in electrical engineering.) The pressure difference $\Delta p$

$$p_C := p(V_T) - p_B = \frac{V_T}{C} \qquad (5)$$

is sometimes called compliance pressure in respiratory text books (which is equivalent to the voltage $\Delta U$ across a capacitor in electrical engineering) and the associated compliance power is:

$$P_C = \frac{V_T^2}{2C} \cdot F \qquad (6)$$

because the work (4) has to be done for each inspiration. With quasi-static spontaneous breathing (due to the conservation of energy theorem) the same physical work respectively power has to be done by the organism, consequently the equations (4) and (6) analogously applies to spontaneous breathing as well.

With (P) it makes sense for the work (4) to assume a minimum under quiet spontaneous breathing which at the same time minimizes the power (6) when the breathing rate $F$ (and due to Eq. (1) $V_T$ as well) is initially given. However, this is only possible if $C$ is as large as biologically possible, or in a mathematical language: The parameter $C$ in Eq. (3) has to be a maximum of the function $\frac{dV}{dp}(p)$ at the point $p = p_B$ which means that $\frac{dC}{dp}(p_B) = 0$ applies and consequently the 2nd derivative $\frac{d^2V}{dp^2}(p_B) = 0$ disappears. Exactly this defines mathematically an inflection point of the function $V(p)$ at the point $p = p_B$ which justifies the above assumption in retrospect and the course of the function $V(p)$ in figure 3 as well as in textbooks (Chatburn 2003; Bryan 2011) and publications (Rahn 1946). This is surely why Fernandez (Fernandez 1993) writes that $C$ is constant at the $FRC$ and Chambers et al. (Chambers 2019) on p. 51 write in their book that lung compliance is at its highest at $FRC$, however, both without citing a reference. On the other hand, this result strenghtens the trust in (P).

The above simple derivation has not been published until now, possibly because it is obvious anyway.

By the way, the resting volume $FRC_{Th}$ of the isolated thorax is larger than the $FRC = V(p_B)$, because the physiologic resting volume of the isolated lungs $FRC_L = 0$ disappears. Nevertheless, quasi-static linear approximations $V_{Th} - FRC_{Th} = (p_P - p_B) \cdot C_{Th}$ and

$$\Delta V_L = (p_A - p_P) \cdot C_L \qquad (7)$$

(the last equation within the range of quiet breathing) similar to (3) (West 2012; Lumb 2016b) are well known, wherein $C_{Th}$ and $C_L$ are the compliances of the isolated thorax respectively isolated lungs. Physiologically the volume of the pleural space is negligible and the volume of the empty lung is much smaller than the $FRC$. Hence, $V_{Th}$, $V_L$ and $V$ are almost equal which is why these volume indices will be suppressed in the following. Often the serial addition equation $\frac{1}{C} = \frac{1}{C_L} + \frac{1}{C_{Th}}$ is quoted in respiratory physiology text books (Levitzky 2018; West 2012) however, this presupposes $FRC_{Th} = FRC = FRC_L$ which is not the case. In an electrical engineering language: The lungs can be compared with an electrolytic capacitor, since only a positive volume [≅ charge] can be applied, however the thorax (like any other capacitor) can be "charged" with a positive or negative volume. Since $FRC_{Th} \gg FRC_L = 0$ the following is true: The lung within the thorax, both together in the resting (= balance) position, unloads the thorax by the volume $(FRC_{Th} - FRC)$ while the thorax keeps the lungs in a loaded position with the volume $FRC$. This is why the pleural pressure $p_P$ has to be negative compared to the ambient pressure $p_B$ in this resting position. As already mentioned, the $FRC$ is the physiologic resting (= balance) position of the lungs (within the thorax) at the end of expiration. If we apply the inverse function $p(V)$ to $V(p_B) = FRC$ this results to $p_B = p(FRC)$. Hence, $p_P < p_B$ holds at the end of expiration. Even more, it is known that $p_P < p_B$ holds during the eintire quiet respiratory period, especially during quiet expiration as well (Hall 2020; West 2012) which is often an argument that quiet expiration is passive.

If ventilation was previously considered to be quasi-static, then this should no longer apply in the following, especially when discussing airway resistance. Consequently $p_A = p_V$ does not apply any more.

## Theoretical considerations on the type of volume flow within the dead space, the alveolar pressure and the respiratory time constant

This and the following subsection are not (or only indirectly) included in the calculation of the main result (29). The volume $V(p_A, t)$ in the previous subsection was mainly a function of the pressure $p_A$, from now on, this volume is mainly a function of the time *t*. Therefore, strictly speaking it would be more appropriate to use the notion of partial derivatives, however, due to convention in physiology, let us exceptionally dispense with the mathematical exact notion. A turbulent flow $\frac{dV}{dt}$ leads to a considerably higher flow resistance which obviously contradicts (P) at least in the resting state. Nevertheless a turbulent resistor summand of the form in Eq. (17) was the basis for the results in previous publication (Otis 1950) dealing with the human respiratory rate. Due to (P), however, it is more likely that the flow within the dead space is laminar, at least for the most part, at rest. This assumption shall now be evaluated and will be shown indirectly in the next subsection using the example of expiration under mechanical ventilation. Then the law of Hagen-Poiseuille applies in each individual dead space section with a cross-section assumed to be nearly circular, because breathing gases (such as air) are Newtonian fluids. It would now be easy to derive the Hagen-Poiseuille law from even simpler physical laws in a few lines, but this should be dispensed with here, since this can be found in physics textbooks. The resistance $R$ of the considered dead space section, bronchus, bronchiolus, trachea, pharynx or nose is therefore dependent on the pipe geometry (length, diameter) as well as the dynamic viscosity of the breathing gas. The pressure difference $p_R$ which arises from the volume flow $\frac{dV}{dt}$ at $R$ is then directly proportional to $\frac{dV}{dt}$:

$$p_R = R \cdot \frac{dV}{dt} \qquad (8)$$

The analogue to (8) in electrical engineering is Ohm's law $U = R \cdot I$. Starting from the alveolar space, the bronchial tree, including pharynx and nose, can be understood as a network (which is rather a tree) of resistors. Analogous to electrical engineering, from the linearity between $p_R$ and $\frac{dV}{dt}$ in each individual dead space section with a resistance $R$ then follows that two serial resistances $R_1$ and $R_2$ must be added $R = R_1 + R_2$ and two parallel resistances $R_1$ and $R_2$ follow the law $\frac{1}{R} = \frac{1}{R_1} + \frac{1}{R_2}$. If these serial and parallel addition rules are applied iteratively throughout the entire resistance network, then the resistance $R$ in Eq. (8) may also be understood as the resistance of the entire airway system which is then usually referred to as airway resistance $R$. This is to be defined in the following, with $p_R$ in Eq. (8) then being the difference between the ambient pressure $p_B$ and the alveolar pressure $p_A(t)$ (figure 4) provided that $p_A(t)$ is a common pressure for all alveoli:

$$p_R(t) = p_B - p_A(t) = R \cdot \frac{dV}{dt} \tag{9}$$

In respiratory physiology this $p_R$ is sometimes called resistance pressure. (Under spontaneous breathing $p_B$ is the constant barometric pressure and under positive pressure ventilation $p_B$ has to be replaced with the seemingly changing ambient pressure $p_V(t) > p_B$ in front of the face respectively lungs.) Of course, during inspiraton the alveolar pressure $p_A$ is smaller then the ambient pressure $p_B$ and during expiration the other way around. Hence, the volume flow $\frac{dV}{dt}$ (as usually defined in respiratory physiology) and the pressure $p_R(t)$ are positive during inspiration and negative during passive expiration.

Admittedly, the geometry of the nose differs largely from a pipe and the flow within the nose might not be laminar. This is possibly why, we usually run with an open mouth. However it is assumed here that the intranasal flow is at least mostly laminar, in any case at rest, since in this state our mouth usually remains closed.

**Definition:** The difference $p_D(t) := p_B - p_P(t)$ between the ambient pressure $p_B$ and pleural pressure $p_P(t)$ is called driving-pressure (s. figure 4).

The driving-pressure $p_D$ is responsible for ventilation under spontaneous breathing and is in some sense the spontaneous ventilation "analogue" to the ventilation pressure $p_V$ during positive pressure mechanical ventilation.

Since the pleural space which is filled with a small amount of serous fluid, surrounds the lungs, the driving pressure $p_D$ is an (at least largely) common pressure for all alveoli. In the following it is assumed that $p_D$ is a common pressure for all alveoli.

At the internal end of the resistance network the alveoli are located. If we compare the lungs to a tree, then the branches correspond to the bronchial tree and the leaves to the alveoli. From a distance, trees of the same species look similar, but the leaves (like the alveoli) are still differently shaped and sized. It is unlikely that physiology would (modulo gravitational effects) preferentially ventilate individual alveoli, thus disadvantage others. Hence, from a macroscopic point of view, the lung is almost homogeneous. Concerning ventilation, the units (alveoli, resistors) of the lungs are large enough, so that quantum effects should only play a minor role. Hence, let us assume the following postulate:

**Postulate:** Each lung segment (e.g. alveolus with associated resistor) follows the same macroscopic physical equations (7) and (8) in any case lastly within the eintire lungs.

Thereof doesn't already follow that the product $R \cdot C_L$ of the parameters in the equations (7) and (8) is homogeneous throughout the lungs, since the alveoli are different in size and shape and in addition the resistance tree is of complex structure.

At first, let us assume that the resistors $R$ and the compliances $C_L$ in Eq. (7) resp. (8) are constant. This assumption surely is justified during quiet breathing, but not under physical exercise. Hence, this assumption shall be corrected at the end of this subsection.
Although the equations (7) and (8) might be valid only within the association of the entire lungs, nevertheless, let us build up the lungs step by step from its alveoli and its resistors by using these two equations.

First of all, let us assume that the lungs consisted of only 2 alveoli with compliances $C_1$ and $C_2$ which are each parallel connected to the internal end of the trachea via the resistors $R_1$ and $R_2$. Let us further assume that the alveolar pressure is not a common pressure within these 2 alveoli, then at the beginning and at the end of each inspiration or expiration (i.e. at the time when the volume-flow at the outer end of the nose equals zero) pressure equalization between these 2 alveoli would take place. This of course additionally requires breathing work which contradicts (P) and simultaneously raises the following question: Which structure of the lungs is necessary, to prevent this additional breathing work, of course during physical exertion as well and consequently for every respiratory rate and breathing pattern? Pressure equalization between alveoli does not take place, if the pressure within the alveoli is always the same. Since the lung itself does not have any information about the very next following respiratory rate, this specifies the above question: Which structure of the lungs is necessary, to keep a common pressure (for all driving pressure patterns) within the alveoli?

A sufficient condition therefore surely is that the compliances and resistances are all equal. Consequently, let us assume that the lungs consisted again of only 2 alveoli, however with the compliances $C_1 = C_2$ which are each parallel connected to the outer end of the nose via the 2 resistors $R_1 = R_2$. Since the driving-pressure $p_D$ for these two alveoli is always the same, these 2 alveoli are ventilated synchronously and due to symmetry there cannot be any alveolar pressure difference $\Delta p_A$ between them. Because the alveolar pressure $p_A$ is the pressure "between" these $C_i$ and $R_i$ $i \in \{1,2\}$, the total compliance $C_L = C_1 + C_2$ of this 2-alveoli-sample lung and the total inverse resistance $\frac{1}{R} = \frac{1}{R_1} + \frac{1}{R_2}$ follow from the symmetry as well.

The analogy between respiratory physiology and electrical engineering supports the last conclusion. (s. table 1).

**Definition:** The product of any laminar-flow resistance $R$ (the one in Eq. (8)) connected in series to an alveolus with the compliance $C$ has the physical unit of a time which is why $\tau = R \cdot C$ is called time constant.

The parameter $\tau$ is equivalent to the time constant in electrical engineering.

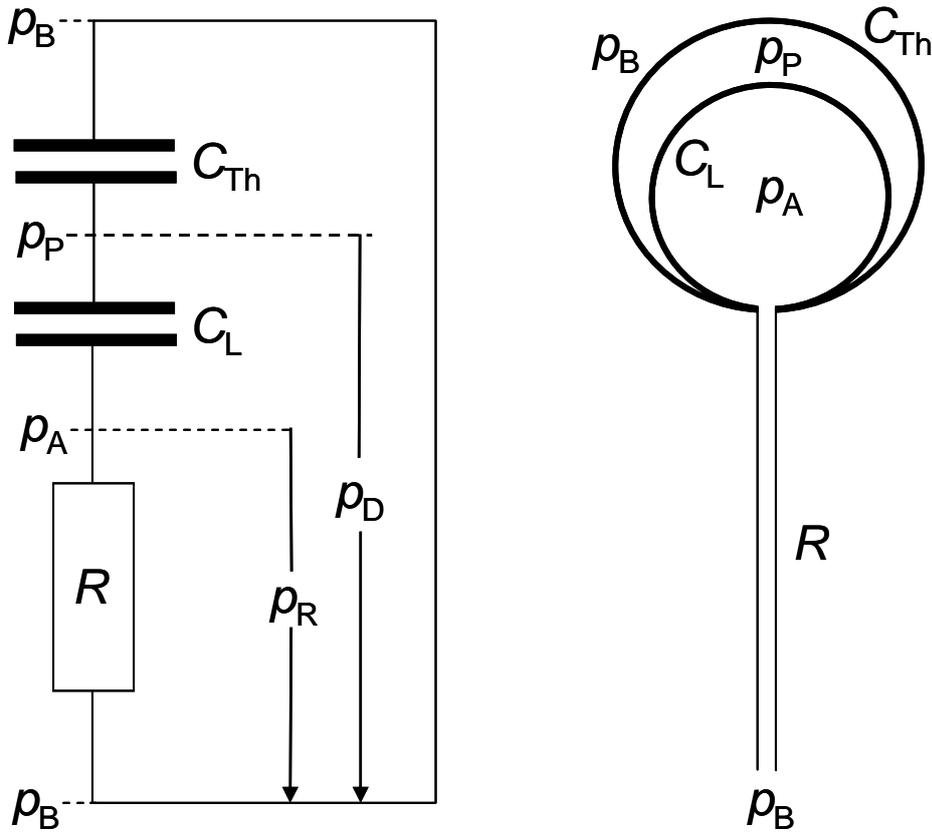

**Figure 4:** Simplified respiratory "circuit diagram": $R$ is the airway resistance, $C$ is the compliance of the lungs respectively thorax and $p$ is the pressure variable. The index P stands for pleural, A for alveolar, B for barometric (or base), R for airway resistance, L for lung, Th for thorax and D for driving pressure. The pleural pressure $p_P$ (and hence $p_D$ as well) is a common pressure for all alveoli. In reality there is almost no space between the lungs and the thorax.

**Claim:** Parallel connection of any alveoli with the same time constant $\tau$ does not change the time constant $\tau$.

(Symmetry)-proof: For the above symmetric 2-alveoli (connected to the outer end of the nose) sample lung with $R_1 \cdot C_1 = R_2 \cdot C_2 = \tau$ the time constant of this small lung is $C_L \cdot R = (C_1 + C_2) / \left( \dfrac{1}{R_1} + \dfrac{1}{R_2} \right) = \tau$. The same proof applies for any ($n \in \mathbb{N}$) $n$-alveoli $C_1 = ... = C_n$ sample lung, all parallel connected to the outer end of the nose with their resistor $R_i = \dfrac{\tau}{C_i}$ $\forall i \in \{1...n\}$. Consequently (apart from units) again for a 2-alveoli sample lung with compliances $C_i = \dfrac{m_i}{n_i}$ $m_i, n_i \in \mathbb{N}; i \in \{1, 2\}$ and thus $R_i = \dfrac{n_i}{m_i} \tau$ with rational numbers (which is sufficient for physiology), since: $C_L = \dfrac{m_1}{n_1} + \dfrac{m_2}{n_2} = \dfrac{1}{n_1 n_2} (m_1 n_2 + m_2 n_1) = \dfrac{\tau}{R_1} + \dfrac{\tau}{R_2} = \dfrac{\tau}{R}$. This again iteratively applies to any number $n \in \mathbb{N}$ of any alveoli $C_i \in \mathbb{Q}^*_+$ $i \in \{1,...,n\}$ each with the same time constant $\tau > 0$ connected in parallel with the resistor $R_i = \dfrac{\tau}{C_i}$ to the outer end of the nose.

**Definition:** In the following an RC-element $(R,C)$ is defined to be the connection of one alveolus with the compliance $C$ to its resistor $R$ or equivalently in electrical engineering the serial connection of one capacitor $C$ to one resistor $R$.

The restriction to rational numbers in the above proof is not necessary: Because of the analogy between electrical engineering and respiratory physiology (s. table 1) the complex impedance calculus in electrical engineering (Quade 1937) can be applied here as well. But even more:

**Claim:** Two parallel connected RC-elements ($R_1, C_1$ respectively $R_2, C_2$) are equivalent to the one RC-element ($C_L = C_1 + C_2$, $R = 1/\left(\dfrac{1}{R_1} + \dfrac{1}{R_2}\right)$) if and only if the time constants of the 2 parallel connected RC-elements match.

(Equivalent in the following sense: From a black box view these behave equivalently if sine-waves of any frequency $\omega$ are applied to them.)

(Complex impedance calculus)-proof: Let us begin with the easier direction: Hence, let $C_1 \cdot R_1 = C_2 \cdot R_2 = \tau > 0$ and let $\omega \in \mathbb{R}^*$ be arbitrary, then a simple calculation shows that the following complex calculus equation holds: $\sum_{i=1}^{2} \dfrac{1}{\dfrac{1}{\omega j C_i} + R_i} = \dfrac{1}{\dfrac{1}{\omega j C_L} + R}$. Thereof $R \cdot C_L = \tau$ follows as well.

Now the proof into the other direction: Let $C_1, R_1, C_2, R_2, R, C_L \in \mathbb{R}_+^*$, then

$$\sum_{i=1}^{2} \dfrac{1}{\dfrac{1}{\omega j C_i} + R_i} = \dfrac{1}{\dfrac{1}{\omega j C_L} + R}$$ holds $\forall \omega \in \mathbb{R}^*$ only if $C_1 \cdot R_1 = C_2 \cdot R_2$, since: The imaginary part

$$\dfrac{\dfrac{1}{C_1}}{R_1^2 + \dfrac{1}{\omega^2 C_1^2}} + \dfrac{\dfrac{1}{C_2}}{R_2^2 + \dfrac{1}{\omega^2 C_2^2}} = \dfrac{\dfrac{1}{C_L}}{R^2 + \dfrac{1}{\omega^2 C_L^2}}$$ and the real part

$$\dfrac{R_1}{R_1^2 + \dfrac{1}{\omega^2 C_1^2}} + \dfrac{R_2}{R_2^2 + \dfrac{1}{\omega^2 C_2^2}} = \dfrac{R}{R^2 + \dfrac{1}{\omega^2 C_L^2}}$$ have to hold separately. After bringing each equation to a common divisor the polynomials in the variable $\omega$ in the nominator have to match on both sides of the equal sign for each $\omega^n$-part with $n \in \{0, 2, 4\}$. Thereof $C_1 \cdot R_1 = C_2 \cdot R_2 =: \tau$, $C_L = C_1 + C_2$, $\dfrac{1}{R} = \dfrac{1}{R_1} + \dfrac{1}{R_2}$ and of course $\tau = C_L \cdot R$ as well follows.

This calculus is called complex, since $j := \sqrt{-1}$ occurs in the calculation and the parameter $\omega$ is called sinusoidal angular frequency. This proof is admittedly a little more general, but hides the symmetry in the previous proof.

**Claim:** Let $k \in \mathbb{N}$, then $k$ parallel connected RC-elements $(C_i, R_i)$, $i \in \{1,..,k\}$ are equivalent to the one RC-element ($C_L = \sum_{i=1}^{k} C_i$, $R = 1/\sum_{i=1}^{k} \frac{1}{R_i}$) if and only if the time constants of all the $k$ parallel connected RC-elements match (figure 5).

Proof: This follows from the above claim by induction from $n-1$ to $n$.

**Definition:** Let us from now on call this $(C_L, R)$ RC-element the equivalent one to the parallel connection of these RC-elements $(C_i, R_i)$ $i \in \{1,..,k\}$.

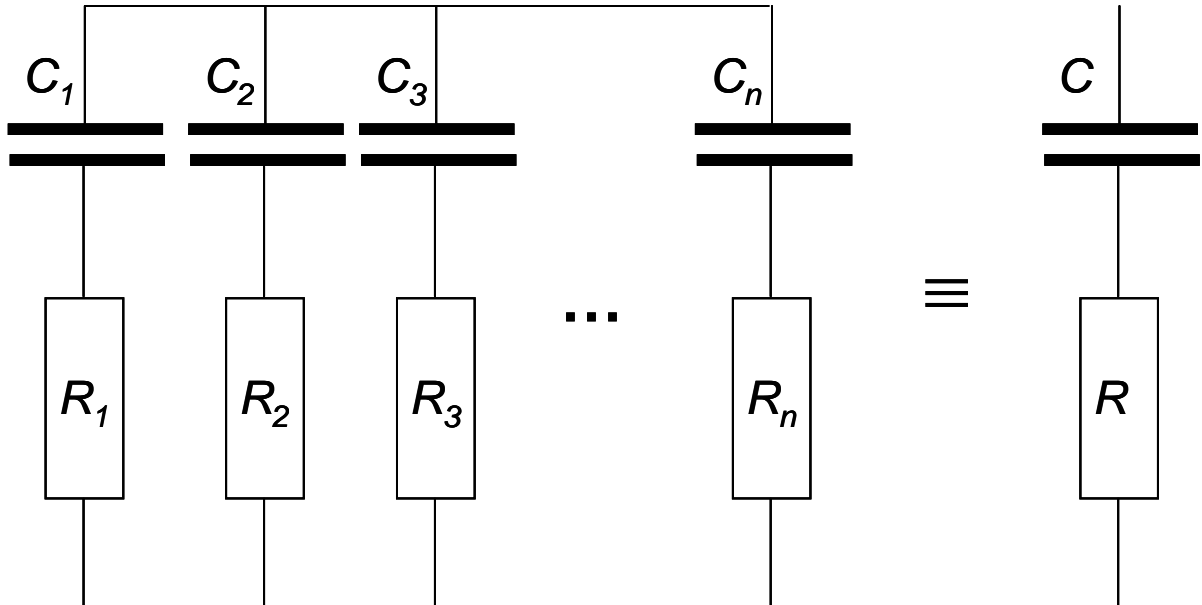

**Figure 5:** The parallel connected RC-elements behave equivalently to the one RC-element (with $C = \sum_{i=1}^{n} C_i$ and $\frac{1}{R} = \sum_{i=1}^{n} \frac{1}{R_i}$) if and only if the time constants $R_i \cdot C_i$ of all the parallel connected RC-elements match. Hence, a lung with many alveoli, all parallel connected with their resistor to the outer end of the nose behaves equivalently to a lung with one alveolus with $C = \sum_{i=1}^{n} C_i$ connected with its resistor $R = 1/\sum_{i=1}^{n} \frac{1}{R_i}$ to the outer end of the nose if and only if all have the same time constant.

If a sine wave of any frequency $\omega$ is applied to parallel connected RC-elements, this simple "lung" behaves equivalently to the one equivalent RC-element (= even a simpler lung with only one alveolus connected to its resistor) if and only if all the parallel connected RC-elements have the same time constant $\tau$.

Let $p_D(t) \in C^2$ in the following be any twice continuous differentiable breathing pattern within $[0, T_I + T_E]$. It is well known that Fourier series converge uniformly to $p_D(t)$. From electrical engineering it is well known that differential equations of RC-elements are linear, consequently the superposition principle holds.

**Corollary:** Parallel connected RC-elements behave for all these $p_D(t) \in C^2$ equivalently to the one equivalent RC-element if and only if all the parallel connected RC-elements have the same time constant $\tau$.

Proof: From a theoretical (or better electro-technical) point of view, all sine (and cosine) functions are a subset of $p_D(t) \in C^2$. The other direction follows from the superposition principle, since the Fourier series of any $p_D(t) \in C^2$ (within any compact interval of the finite live) is composed of sine (and cosine) functions. (Because of the above mentioned uniform convergence, a finite Fourier series sum surely is sufficient for biology).

By the way, from a biological point of view, the condition $p_D(t) \in C^2$ is practically always fulfilled, since (due to the Stone-Weierstrass theorem) every continuous function defined on a closed interval can be uniformly approximated as closely as desired by a polynomial function and polynomials are even analytic.

**Corollary:** Let $k \in \mathbb{N}$, then $k$ parallel RC-elements behave for all $p_D(t) \in C^2$ equivalently to the one equivalent RC-element if and only if all RC-elements have the same time constant. (The one equivalent RC-element of course again has this time constant.)

**Corollary:** The alveolar pressure $p_A(t)$ within a *2*-alveoli (each alveolus connected by its resistor to the outer end of the nose) sample lung is a common pressure for all $p_D(t) \in C^2$ only if the two alveoli have the same time constant.

Proof: (Of course these *2*-alveoli behave like 2 parallel connected RC-elements.) This follows from the above symmetry proof or the fact that this *2*-alveoli sample lung is equivalent to the equivalent 1-alveolus sample lung and the pressure within this last one is by definition a common pressure.

If the time constants differ: $C_1 \cdot R_1 \neq C_2 \cdot R_2$, then there exists a breathing pattern $p_D(t) \in C^2$, so that the alveolar pressure still remains a common pressure, namely $p_D(t) = 0 \; \forall t \in \mathbb{R}$ (hence, $\forall t \in [0, T_I + T_E]$ above all) which is of course biologically irrelevant (and is an exceptional case anyway). The elaborate proof that this is an exceptional case shall be dispensed here, since this result is not necessary in the following. Nevertheless let us study the following example:

**Claim:** If the time constants differ, then there exist breathing patterns, so that the alveolar pressure within the above 2-alveoli sample lung is not always a common pressure.

Proof: Let $C_1 \cdot R_1 \neq C_2 \cdot R_2$ and the lungs shall consist of the 2 alveoli with the compliances $C_1$ and $C_2$ which are each parallel connected to the outer end of the nose via their resistors $R_1$ and $R_2$. The analogy between electrical engineering and respiratory physiology (s. table 1) shows that during passive expiration the following alveolar pressure difference $\Delta p_A(t) = p_1 \cdot e^{-t/R_1 C_1} - p_2 e^{-t/R_2 C_2}$ for all $t \geq 0$ applies, whereby $p_1$ and $p_2$ are the alveolar pressures within these alveoli at time $t = 0$. This pressure difference $\Delta p_A(t)$ disappears $\forall t \geq 0$ of course only if $C_1 \cdot R_1 = C_2 \cdot R_2$, since the right hand side of $\dfrac{p_1}{p_2} = e^{t(1/R_1 C_1 + 1/R_2 C_2)}$ is a function of the time *t*, whereas the left hand side is constant.

Hence, a common time constant is a necessary condition to keep a common alveolar pressure within this small sample lung.

**Corollary:** The alveolar pressure $p_A(t)$ is not a common pressure for all $p_D(t) \in C^2$ of the above *n*-alveoli (connected to the outer end of the nose) sample lung if (only) one time constant differs from all the other ones.
Proof: Collect all the $n-1$ alveoli with the same time constant to 1 equivalent RC-lung (better sub-lung). This corollary then follows from the last one.

**Corollary:** The alveolar pressure $p_A$ of a *n*-alveoli (each connected to the outer end of the nose) sample lung is a common pressure for all alveoli and all $p_D(t) \in C^2$, only if all alveoli have the same time constant.

**Claim:** Provided that the this sample lung consists of $n \in \mathbb{N}$ alveoli, each parallel connected to the outer end of the nose via their resistors, all having the same time constant, then this lung is ventilated synchronously.

Proof: Formally this *n* alveoli sample lung is equivalent to a 1-alveoli sample lung, hence this again follows from symmetry.

This is provisionally the above mentioned dimensioning condition for this very small and simple "bronchial tree".

Now back to biology: In reality, the alveoli are jointly ventilated via the resistance tree (bronchi, bronchioles, trachea and nose), but this must not change the time constants $\tau$ and the common alveolar pressure, because:

Let us again assume that the alveolar pressure is not common within all the alveoli, then again at the beginning and at the end of each inspiration or expiration pressure equalization between the alveoli would take place. This additionally requires breathing work which again contradicts (P). Consequently the direction of the volume flow has to be the same within the eintire resistance tree. This direction of course changes between the ventilation phases inspiration respectively expiration. The joint ventilation throughout this resistance tree is then of course as well responsible for the synchronous ventilation and is therefore a dimensioning-condition for the resistance tree as well.

This dimensioning-condition shall now be discussed.

**Claim:** Let us assume that the lungs again consisted of only 2 alveoli with the compliances $C_1$ and $C_2$ which are each parallel connected via their resistors $R_1$ and $R_2$, however this time not to the (outer end of the) nose but to the (internal end of the) trachea with a resistor $R_{\text{common}}$. Then the time constants $R_1 \cdot C_1 = R_2 \cdot C_2$ are still equal and the alveolar pressure is still a common pressure within these 2 alveoli.
Proof: The proof is exactly the same, as before. The only difference is: The (common) ambient pressure $p_B$ (at the nose) has to be replaced with the (common) pressure $p_{common}(t)$ at the internal end of $R_{common}$ which is the connection point to $R_1$ and $R_2$. More precisely: The former driving pressure is now $p_D(t) = p_P(t) - p_{common}(t)$ (better: sub-driving pressure). Consequently $C_1 \cdot R_1 = C_2 \cdot R_2$ and formally the $R_i$ and $C_i$ $i \in \{1,2\}$ can be replaced with $C_L = C_1 + C_2$ and $R = 1/\left(\dfrac{1}{R_1} + \dfrac{1}{R_2}\right)$ whereby $R \cdot C_L = R_1 \cdot C_1 = R_2 \cdot C_2$ holds. The serial connection of $R_{common}$ to this 2 alveoli sample lung (which is formally equivalent to the 1-alveolus sample lung with the equivalent $R \cdot C_L$-element) does not affect the fact that $p_{common}(t)$ remains a common pressure. From the viewpoint at the outer end of $R_{common}$, the time constants increase a little to $(R + R_{common}) \cdot C_L$, but it is physically evident that (due to symmetry) these 2 time constants and consequently $p_A(t)$ still remain equal.

This proof does not change, if more than 2 alveoli (e.g. $n \in \mathbb{N}$ alveoli) each parallel connected via their resistors are (at the same branching point) connected to $R_{common}$. In reality this is often the case. Consequently, each of them again has to have the same time constant and again this small lung can be formally replaced by the equivalent 1-alveolus lung with the same time constant $R \cdot C_L = \left(\sum_{i=1}^{n} C_i\right) / \sum_{i=1}^{n} \dfrac{1}{R_i}$. From the viewpoint at the outer end of $R_{common}$, the time constants again increase a little to $(R + R_{common}) \cdot C_L$.

The eintire lung can now be composed from this small bronchial tree by iterating the above proof (at each branching point) throughout the eintire real resistance tree. The above resistor $R_{common}$ is the parameter which needs then to be adapted at each step. Consequently the dimensioning-condition for the eintire bronchial tree is ultimately a consequence of (P) as well.
Of course, the time constant of the eintire lung is larger, than the time constant of each isolated lung segment (in the broadest and not anatomical sense). The resistance of the trachea, pharynx and nose is then the last time constant increasing step.

**Corollary:** The alveolar pressure $p_A(t)$ is a common pressure within the lungs only if all alveoli have the same time constant.

The following few lines are a little bit "hypothetic" and are not at all necessary for the results of this paper. In all the proofs of this subsection all occurring $C$ and $R$ were assumed to be independent of the volume $V_{Lung}$ within the lungs which approximately is true for small tidal volumes $V_T$ like the ones at rest. However, like $C(p)$ (s. figure 3), $C_L$ is not constant, but a function $C_L(p_{Lung})$ of the pressure $p_{Lung}$ within the isolated lungs (West 2012; Lumb 2016b). By using the same arguments as in the previous subsection, the inverse function $p_{Lung} = p_{Lung}(V_{Lung})$ exists as well. Hence, $C_L$ is also a function $C_L(V_{Lung})$ of the volume $V_{Lung}$ within the lungs. It goes without saying that $C_L(p_{Lung})$ and $C_L(V_{Lung})$ are different functions, nevertheless let us denote them with the same letter as long as there is no risk of confusion. Moreover, it is known that at least close to the residual volume $RV$ or the total lung capacity $TLC$ the airway resistance $R = R(V_{Lung})$ depends also on the intrapulmonary volume $V_{Lung}$ (West 2012). Hence, the equations (7) and (8) for tidal volumes $V_T$ large enough (e.g during physical exertion) have to be adapted in the following way:

$$\Delta V = (p_A - p_P) \cdot C_L(V_{Lung})$$

$$p_R = R(V_{Lung}) \cdot \dfrac{dV}{dt}$$

The postulated homogeneity must not change with the volume $V_{Lung}$ within the lungs. Hence, the parameter $V_{Lung}$ has to be the same for each lung segment (in the broadest and not anatomical sense). Otherwise, homogeneity would be broken. Of course, the functions $C_L(V_{Lung})$ respectively $R(V_{Lung})$ (apart from the physical unit) need not to be the same. Again, thereof doesn't already follow that $R(V_{Lung}) \cdot C(V_{Lung})$ is a homogeneous function for all alveoli, since as mentioned above, these are different in size and shape and in addition the resistance tree is of complex structure. Hence, all the $C$, $C_i$, $C_L$, $R_i$, $R$ within this subsection have to be replaced with $C(V_{Lung})$, $C_i(V_{Lung})$, $C_L(V_{Lung})$ respective $R_i(V_{Lung})$, $R(V_{Lung})$. This presupposes that the serial and parallel addition law's of compliances respectively resistors are still valid. But this trivially is the case and the proof of them changes only slightly. Since the charge $Q$ is the electrical engineering analogue to the volume V, all resistors and capacitors now homogeneously depend on $Q_{Lung}$. For example the parallel addition law of these $Q_{Lung}$-dependent resistors $R_1(Q_{Lung})$ and $R_2(Q_{Lung})$: Due to Ohm's law $U = R_1(Q_{Lung}) \cdot I_1(Q_{Lung}) = R_2(Q_{Lung}) \cdot I_2(Q_{Lung})$ and due to Kirchhoff's current law $I(Q_{Lung}) = I_1(Q_{Lung}) + I_2(Q_{Lung})$ holds, whereof $R(Q_{Lung}) := \frac{U}{I(Q_{Lung})} = \frac{U}{I_1(Q_{Lung}) + I_2(Q_{Lung})} = \frac{U}{\frac{U}{R_1(Q_{Lung})} + \frac{U}{R_2(Q_{Lung})}} = \frac{1}{\frac{1}{R_1(Q_{Lung})} + \frac{1}{R_2(Q_{Lung})}}$ follows. Hence, $\frac{1}{R(Q_{Lung})} = \frac{1}{R_1(Q_{Lung})} + \frac{1}{R_2(Q_{Lung})}$ is true. The proof doesn't even change if $U$ is replaced with $U(Q_{Lung})$. The serial addition law of homogeneously $Q_{Lung}$-dependent capacitors $C_1(Q_{Lung})$ and $C_2(Q_{Lung})$ gives the analogous result as well: Denote $q$ the charge (of course) on both capacitors, then $q = U_1(Q_{Lung}) \cdot C_1(Q_{Lung}) = U_2(Q_{Lung}) \cdot C_2(Q_{Lung})$ and $U(Q_{Lung}) = U_1(Q_{Lung}) + U_2(Q_{Lung})$, hence $C(Q_{Lung}) := \frac{q}{U(Q_{Lung})} = \frac{q}{U_1(Q_{Lung}) + U_2(Q_{Lung})} = \frac{q}{\frac{q}{C_1(Q_{Lung})} + \frac{q}{C_2(Q_{Lung})}}$. Consequently $\frac{1}{C(Q_{Lung})} = \frac{1}{C_1(Q_{Lung})} + \frac{1}{C_2(Q_{Lung})}$ follows. Again, the proof doesn't change if $q$ is replaced with $q(Q_{Lung})$ or $\Delta q$. The proofs of the other serial and parallel addition laws are even simpler. The proofs of $\frac{1}{R(V_{Lung})} = \frac{1}{R_1(V_{Lung})} + \frac{1}{R_2(V_{Lung})}$, $\frac{1}{C(V_{Lung})} = \frac{1}{C_1(V_{Lung})} + \frac{1}{C_2(V_{Lung})}$ and the other serial and parallel addition laws for respiratory physiology are of course equivalent to the ones in this obscure electrical engineering. The proof of the impedance calculus is a little more elaborate, but can be done in the same way. Only the proof using the passive expiration can be applied within a small neighbourhood of any point of $V_{Lung}$, but this is sufficient.

**Corollary:** The "time constant" $R(V) \cdot C_L(V)$ of the isolated lungs is a homogeneous function of the intrapulmonary volume $V$ i.e. all the alveoli (and the lung itself) obey the same "time constant" function.

Otherwise during physical exertion the alveolar pressure would not be a common pressure within all the alveoli and moreover, would contradict (P) in this activity state. You might possibly still doubt this corollary, however, no physiologist doubts the fact, that the alveolar pressure is a common pressure within the lungs and this fact is only possible, if the time constant (function) is homogeneous througout the lungs. On the other hand, interestingly, the term time constant (function) was never defined or even mentioned in any scientific paper until now. However, the present paper deals only with the resting spontaneous breathing, hence this corollary is only a side result and from now on the time constant remains constant again within this paper. Nevertheless, this corollary could partially explain the course of spirometry functions in medicine. This corollary has of course to be biologically and clinically evaluated first which ends the "hypothetic" lines.

From a mathematical point of view, the time constant of all the alveoli within the lungs might possibly not be exactly the same. From a biological point of view, despite the fact that the pleural pressure $p_P(t)$ is a common pressure for the lungs, small differences in the driving pressure $p_D(t)$ might lead to small differences in the alveolar pressure $p_A(t)$. However, the elastic septa between the bronchioles and in particular between the alveoli are so thin (Peake 2015) that pressure equalization of small alveolar pressure differences takes place almost immediately.

Consequently, ultimately following from (P), it can be concluded that the alveolar pressure $p_A(t)$ is an (at least largely) common pressure within real lungs and all alveoli have (almost) the same time constant (or "hypothetic time constant" function). This is now the answer to the above given question.

The next question arises now: How can the laminar flow assumption (8) be proven?

**Claim:** The passive (laminar-flow) expiration function of an 1-alveolus sample lung with the constant compliance $C_L$ connected to a resistor $R$ is the following exponential function:
$$V(t) = V(0) \cdot e^{-t/R \cdot C_L} \tag{10}$$
Proof: The analogy between pulmonary physiology and electrical engineering shown in table 1 gives this result. $V(0)$ is of course the volume within the alveolus at the beginning of expiration. The proof will be dispensed here, since the analogue proof can be found in every electric engineering textbook and the next subsection derives this proof for the lung within the thorax anyway.

During quiet spontaneous breathing the compliance $C_L$ is almost constant within a neighbourhood of the $FRC$ and the constancy of $R$ within the quiet breathing range will be discussed in the next subsection. Hence, due to the derived homogeneity of $\tau = R \cdot C_L$ together with the common alveolar pressure $p_A$ the following holds:

**Corollary:** The passive quiet expiration function of any real (isolated) lungs in biology is an exponential function (10).

However, no lung is isolated during spontaneous breathing. Due to Eq. (3) within a neighbourhood of the *FRC* the compliance $C$ is almost constant as well and during quiet spontaneous breathing the tidal volume $V_T$ remains within the constant range of $C$. The serial connection of the thorax compliance $C_{Th}$ to the lung compliance $C_L$ surely changes the time constant of all the alveoli, however the time constants of all the alveoli still remain equal. Hence the constant $C_L$ in (10) can be replaced with the constant $C$ if $V_T$ remains small enough. This surely is the case during passive quiet spontaneous expiration. Note: This argumentation does not need the above mentioned compliance serial addition equation $\frac{1}{C} = \frac{1}{C_L} + \frac{1}{C_{Th}}$. The electrical engineering argument would be: We just replace one capacitor by another one.

Consequently, the time constant of the lungs within the thorax during spontaneous breathing is $C \cdot R$. This result can now be used as a test for the laminar flow assumption (8) which will be the subject of the next subsection.

Nevertheless, there are gravitational physiologic and (patho)physiologic factors (e.g. mucus) influencing the airway resistance or compliance which is probably why nature has invented the bronchial muscles for fine adaptation of the individual alveoli time constants. Moreover, since $C_L(p)$ is slightly different for inspiration and expiration (West 2012), this fine adaptation should depend on the direction of the volume flow. The last seems to be clinically unproven until now and should be evaluated in the future. Disturbances, like asthma bronchiale seem to support the existence of a flow direction dependent fine adaptation.

Interestingly, these theoretical considerations concerning the time constant, synchronous ventilation and common alveolar pressure (due to (P)) were not published until now. Possibly this is why, the results of equation (17) are still regarded to be valid and consequently the calculation of the respiratory rate at rest (Otis 1950) were not reevaluated until now.

## *Passive expiratory volume-function and flow-pattern of the lungs*

In the following, the time course of the volume $V(t)$ and the flow pattern $\frac{dV}{dt}$ during quiet passive expiration is to be clarified. The theoretical considerations in the previous subsection led (due to (P)) to the fact that the entire lung within the thorax has a (largely) unique time constant and the alveolar pressure $p_A$ is a common pressure for all alveoli. This theoretical result shall now be evaluated. As mentioned above, expiration is passive both during spontaneous breathing and controlled ventilation, the energy required for this being obtained from (4). Mechanical controlled ventilation is sometimes performed under muscle relaxation. Hence, as a test for equation (10), controlled ventilation (with regard to spontaneous breathing exceptionally without *PEEP*) is better suited. The electrical engineering analogy to the passive expiration is the unloading of a capacitor which leads due to Eq. (3): $p - p_B = \frac{(V - FRC)}{C}$ together with Eq. (9) to the first order homogeneous linear differential equation $\frac{(V - FRC)}{C} + R \cdot \frac{dV}{dt} = 0$. Another argument, leading to this equation is the application of Kirchhoff's voltage law which corresponds to the Kirchhoff's pressure law in respiratory physiology. After separation of variables this simple differntial equation gives:

$$\int \frac{dV}{(V - FRC)} = \int \frac{-dt}{R \cdot C} + Const \quad (11)$$

$\Rightarrow \ln(V - FRC) = \frac{-t}{R \cdot C} + Const. \Rightarrow V(t) = FRC + e^{Const.} \cdot e^{-t/\tau}$ , wherein $\tau = R \cdot C$ is the time constant of the lungs within the thorax. Based on the initial condition $V(0) = FRC + V_T$ which corresponds to the intrapulmonary volume at the beginning of expiration $e^{Const.} = V_T$ follows and for sufficiently large $t$ (since ventilation was performed without $PEEP$) the $FRC$ is the "asymptote" of $V(t)$ until the next inspiration. As expected, the time course of the intrapulmonary volume during expiration is an exponential one:

$$V(t) = FRC + V_T \cdot e^{-t/\tau} \quad (12)$$

Since the $FRC$ ultimately remains within the lungs under mechanical ventilation, the function $V_T \cdot e^{-t/\tau}$ is specified in textbooks (Davies 2005; Chatburn 2003), publications (Bergman 1969; Botsis 2003; Ashutosh 1978) and also shown on modern ventilators as the passive expiration curve (figure 6). This result was even already published by Brody in 1954 (Brody 1954).

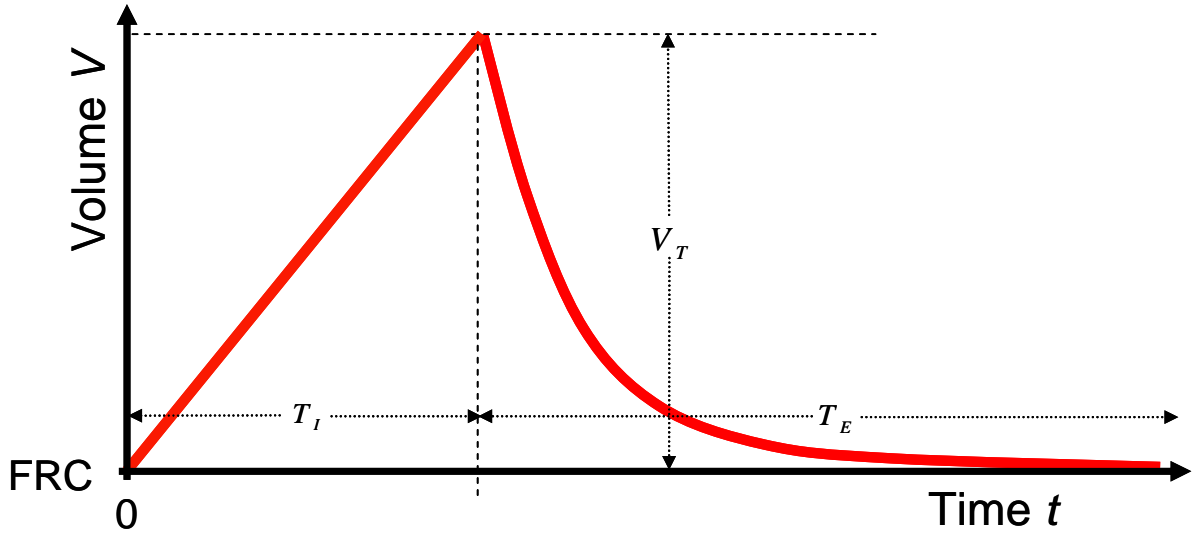

**Figure 6:** After inspiration with a constant volume flow starting from the intrapulmonary volume $FRC$ to the volume $FRC + V_T$, expiration takes place passively with the time constant $\tau = R \cdot C$ under ventilation. The $FRC$ is the "asymptote" of the expiratory function $V(t) = FRC + V_T \cdot e^{-t/\tau}$ and is almost reached again after about $5\tau$ because $e^{-5} \approx 0$. This volume-controlled, flow-constant ventilation serves as an idealized model for spontaneous quiet breathing. The quotient $T_I / T_E$ is called $I:E$-ratio.

The time constant $\tau$ was recorded on adult animals (Robinson 1978) and on several newborn mammals (Mortola 1983; Mortola 1985) as well which is why the result (12) of course is not human-specific. Completely analogue to the unloading of a capacitor, about $e^0 - e^{-1} \approx 63\%$ of the tidal volume $V_T$ escaped from the lungs after one $\tau$ and after $5 \cdot \tau$ the $FRC$ is almost reached again because $e^0 - e^{-5} \approx 99.3\%$. These two percentages are published in respiratory medicine textbooks often without citing Eq. (11) (Cairo 2012).
From Eq. (12) the volume flow pattern:

$$\frac{dV}{dt} = -\frac{V_T}{\tau} \cdot e^{-t/\tau} \tag{13}$$

which usually is also monitored on modern ventilators and with Eq. (5) the pressure gradient (9):

$$p_B - p_A(t) = \frac{-R \cdot V_T}{\tau} \cdot e^{-t/\tau} = -p_C \cdot e^{-t/\tau} \tag{14}$$

follow an exponential function as well. The signs indicate the direction of flow and pressure gradient.

The product of the volume flow $\frac{dV}{dt}$ and the pressure difference $\Delta p$ which this $\frac{dV}{dt}$ causes across a resistor $R$ has the physical unit of a power. The time integral of the power $\frac{dV}{dt}(p_B - p_A(t)) = \frac{dV}{dt} \cdot p_R(t)$ is therefore the work $\left| \int_0^\infty \frac{dV}{dt} \cdot p_R(t) \, dt \right|$ which is passively done on the airway resistance $R$ during expiration and ultimately converted into heat. As you can see for yourself by inserting Eq. (13) and Eq. (14), this work corresponds exactly to (4).

Let $T_E$ denote the expiratory duration and $T_I$ the inspiratory duration, then e.g. the data of adult humans from table 2 show that

$$T_E / \tau \approx 10 \text{ and } T_I / \tau \approx 5 \tag{15}$$

Because of $\int_{T_E}^\infty e^{-2\frac{t}{\tau}} dt = \frac{\tau}{2} e^{-2\frac{T_E}{\tau}} \approx \tau \cdot 10^{-9}$, the upper limit $\infty$ in the above integral can be replaced by $T_E$ with a negligible error. By the way, the data from table 2, show that the quotients (15) for the human newborn are even larger. This last fact shall be discussed later in connection with the $T_I / T_E := I : E$ ratio (s. figure 6).

From (15) together with (14) $p_A(T_E) - p_B = p_C \cdot e^{-T_E/\tau} \approx p_C \cdot e^{-10} \approx 0$ follows. Since the left hand side eqals the definition of the $PEEP$, there is (almost) no physiological $PEEP$, a well known fact. From (15) now together with (13) the quasi-static condition $\frac{dV}{dt}(T_E) = \frac{V_T}{\tau} \cdot e^{-T_E/\tau} \approx \frac{V_T}{\tau} \cdot e^{-10} \approx 0$ at the end of expiration ($t = T_E$) and once again from (15) together with (12) $V(T_E) = FRC + V_T \cdot e^{-T_E/\tau} = FRC + V_T \cdot e^{-10}$ follows. Consequently, (almost) the entire tidal volume has left the lungs at the end of expiration. After applying the inverse function to $p_A(T_E) - p_B \approx 0 \Leftrightarrow p_A(T_E) \approx p_B$ this gives $V(p_A(T_E)) \approx V(p_B)$. Consequently, (with the initially introduced convention) physiologically $FRC = V(p_B)$ applies with a comparatively small error, which justifies the linear approximation (3) in retrospect as well. This procedure was of course allowed due to the quasi-static condition at $t = T_E$. Concerning spontaneous breathing, because of the change in sign (between the breathing phases) of the continuous intrapulmonary pressure curve $p_A(t)$ (with the intermediate value theorem), there must even be exactly one point on $V(p)$ for which $V(p_B) = FRC$ applies.

If the assumption of a laminar flow in the dead space which led to Eq. (8), were wrong, then the expiration equation would not follow an exponential function, because a turbulent flow typically shows a resistance $R_T$ behavior that does not depend linearly but approximately on the square (Lumb 2016c) of the volume flow $\frac{dV}{dt}$:

$$\Delta p = R_T \cdot \left(\frac{dV}{dt}\right)^2 \tag{16}$$

An exponential expiratory behavior as illustrated in figure 6 can never be achieved with this - the expiration would be considerably more slowly. However, since the exponential volume curve (12) is described under ventilation during passive expiration (Davies 2005; Chatburn 2003; Bergman 1969; Botsis 2003; Ashutosh 1978; Mireles-Cabodevila 2020) and is also displayed in this way on ventilators, it can, conversely, be assumed that turbulent flow behavior at rest, if at all, plays a subordinate role at best which justifies the laminar flow assumption and consequently Eq. (8) in retrospect. Of course, this is not a physical but biological proof, nevertheless the well known fact (12) sufficiently supports the previous theoretical considerations which ultimately base on Eq. (8). On the other hand, these results (common alveolar pressure, common time constant) should again strenghten the trust in (P).

Incidentally, with a turbulent flow (16) would have the consequence that the "time constant" $\tau = R_T \cdot C$ has the physical unit $s^2 \cdot m^{-3}$, so the term time constant would no longer be appropriate and if

$$\Delta p = R \cdot \frac{dV}{dt} + R_T \cdot \left(\frac{dV}{dt}\right)^2 \tag{17}$$

were to apply, then $\tau$ could no longer be assigned a clear physical unit. Nevertheless, this equation (17) was used for the calculation of the respiratory rate at rest in previous publications (Otis 1950) and the results thereof are still expected to be valid.

Concerning (11) this step is only allowed, if the time constant $R \cdot C$ is independent of the volume $V$ during quiet breathing. For the compliance $C$ with the linearization (3) at the inflection point of $V(p)$ this surely is the case. The nose, pharynx, throat, trachea and bronchi which are surrounded by bones or cartilage (Peake 2015), are mainly responsible for $R$ (West 2012), while the bronchioles only make a very small contribution to $R$. Consequently, this mainly responsible part for $R$ should not (or only insignificantly) depend on the volume $V$ in the lung, at least during quiet breathing. Nevertheless, if $R$ depends on the volume already for $V \in [FRC, FRC + V_T]$ then an exponential expiratory behavior like Eq. (12) or Eq. (13) could not be achieved. By the way, none of the publications (Otis 1950; Crosfill 1961; Mead 1960; Kerem 1996; Noël 2019) that deal with the resting respiratory rate did even mention this topic.

The dynamic viscosity of the breathing gas is caused by the internal friction of the gas. The internal friction (responsible for the tissue resistance work) within the lung tissue and chest wall structures or between the lungs and thorax usually plays a subordinate role which is why it should either be neglected for the sake of simplicity or could be understood as being subsumed in $R$ (Bates 2015; Bates 1988), as will be suggested here. Assuming this resistance work to overcome the viscosity of the lung and chest wall structures during passive expiration would not show an approximately linear resistance behavior analogous to Eq. (8), this behavior would also disturb the exponential course of (12). However, since (12) is described in the literature and shown in this way on ventilators, it can be assumed, conversely that $R$ in Eq. (8) actually includes all of the above-mentioned friction and that this resistance $R$ also (predominantly) depends linearly (Bates 2015) on $\frac{dV}{dt}$, especially since only Eq. (8) leads to the exponential functions (12), (13) and (14). Moreover, the simple derivation in Eq. (13) would not be valid any more and a deviation from the exponential function would be more obvious in the course of (13).

The quotients (15) are based on human data which is why the objection could be raised that in evolution, humans could be an exception in this regard. But this is very unlikely. Note: Even halving the quotients in (15) would only change the further discussion slightly.

In any case, it would be a waste of energy in the sense of (P) if the compliance work (4) for passive expiration were not recovered analogously to ventilation while breathing spontaneously at rest. It is therefore to be assumed that passive expiration with quiet spontaneous breathing also approximates the exponential function (12). For the sake of simplicity, this should in any case be assumed here. A discussion about this assumption could develop in subsequent publications, but this does not (or at best only insignificantly) change the respiratory power, since quiet expiration is passive (Levitzky 2018; Lumb 2016a).

## *Inspiratory flow pattern and resistance power*

In the following, the course of the volume flow pattern $\frac{dV}{dt}(t)$ during inspiration within $[0, T_I]$ is now to be explored. For biological reasons $\frac{dV}{dt}(t)$ shall be a continuous function. The (positive) pressure gradient $\Delta p = p_R = p_B - p_A$ between the environment and the alveolar space caused by the respiratory muscles during inspiration under spontaneous breathing leads to a (positive) volume flow $\frac{dV}{dt}$ of the breathing gas. The inspiration consequently requires the power $P_R = \frac{dV}{dt} \cdot \Delta p$ to overcome the airway resistance $R$ in Eq. (9). This power $P_R$ is referred to here as the resistance power which is equivalent to $I \cdot \Delta U$ in electrical engineering. With Eq. (9) it follows (s. table 1):

$$P_R = \left(\frac{dV}{dt}\right)^2 \cdot R \tag{18}$$

If the entire time $T_I$ available for inspiration is used for a constant inspiratory volume flow $\frac{dV}{dt}$ which is required to inhale the tidal volume $V_T$ into the lungs, then:

$$\frac{dV}{dt} = \frac{V_T}{T_I} \tag{19}$$

follows. With mechanical ventilation one would describe this as constant flow ventilation. In physiology, as already mentioned above, the $I:E$ breathing time ratio is defined as the quotient $I:E := T_I / T_E$ (s. figure 6) and because of $F = \frac{1}{T_I + T_E}$ the following applies:

$$T_I = \frac{1}{F}\left(\frac{I:E}{I:E+1}\right) \tag{20}$$

as

$$T_E = \frac{1}{F}\left(\frac{1}{I:E+1}\right) \tag{21}$$

First of all, let the respiratory rate $F$ (within a neighbourhood of the resting respiratory rate $F_{opt}$) but also the $I:E$-ratio and thus also $T_I$ due to Eq. (20) be given. With Eq. (1) it follows that with normal quiet spontaneous breathing the tidal volume

$$V_T = V_D + \frac{\dot{V}CO_2}{F \cdot F_{Et}CO_2} \tag{22}$$

is then fixed as well, because (as discussed above) $\dot{V}CO_2$ is largely constant in the resting state. Even if the constant volume flow (19) will ultimately turn out to be the power-minimized one, let us now assume that $\frac{dV}{dt}$ is not constant, but an arbitrary function of time $t$ within the period $[0, T_I]$. Overall, of course, the inspiratory volume flow $\frac{dV}{dt}$ has to fulfill:

$$\int_0^{T_I} \frac{dV}{dt} dt = V(T_I) - V(0) = FRC + V_T - FRC = V_T \qquad (23)$$

The inspiratory resistance power (18) which now has to be averaged (i.e. integrated and divided by the integration interval) over $T_I$, is thereafter:

$$\frac{R}{T_I} \cdot \int_0^{T_I} \left(\frac{dV}{dt}\right)^2 dt \qquad (24)$$

wherein $R$ is given by the anatomy and was therefore (as a constant) pulled in front of the integral. Analogous to the procedure for the compliance power, (24) should now be minimized again because of (P). One could object here that these minimizations might not be independent. However, except $V_T$ and consequently because of Eq. (22) $V_T = V_T(F, const)$ (within a neighbourhood of $F_{opt}$) there are no common parameters in Eq. (6) and Eq. (24). For this reason the "connecting"-parameter $F$ (or $V_T$) requires another use of (P).

The constant volume flow (19) results in the power

$$P_{R, constant\ flow} = \frac{R}{T_I} \cdot \int_0^{T_I} \left(\frac{V_T}{T_I}\right)^2 dt = R \cdot \frac{V_T^2}{T_I^2} \qquad (25)$$

Doubling $2\frac{V_T}{T_I}$ the volume flow $\frac{dV}{dt}$ only needs the time $\frac{T_I}{2}$ to fulfill Eq. (23), but double the power $\frac{R}{T_I} \cdot \int_0^{T_I/2} \left(2\frac{V_T}{T_I}\right)^2 dt = 2R \cdot \frac{V_T^2}{T_I^2}$ which intuitively suggests that $\frac{dV}{dt}$ must fill the entire available time span $[0, T_I]$. What is sought, is a function $\frac{dV}{dt}$ of the time $t$ within $[0, T_I]$ which, under the side condition (23), minimizes the inspiratory resistance power (24). **Claim:** The solution of this variation problem is already (19) with the power (25). Proof: It is now assumed, however that $\frac{dV}{dt} = \frac{V_T}{T_I} + g(t)$ is even better regarding power minimalism, where $g(t) \neq 0$ is any function within $[0, T_I]$ which, because of (23), has to fulfill the condition $\int_0^{T_I} g(t) dt = 0$. The function $g(t)$ of course has to be real and continuous for physiological or biological reasons. With this assumption, the inspiratory resistance power $\frac{R}{T_I} \cdot \int_0^{T_I} \left(\frac{V_T}{T_I} + g(t)\right)^2 dt = \frac{R}{T_I} \int_0^{T_I} \left(\left(\frac{V_T}{T_I}\right)^2 + 2\frac{V_T}{T_I} \cdot g(t) + (g(t))^2\right) dt$ applies. The first summand equals (25), the second one disappears because of the above condition, since $2\frac{V_T}{T_I}$ may be pulled in front of the integral. However, because of $\frac{R}{T_I} > 0$, the third term $\frac{R}{T_I} \cdot \int_0^{T_I} (g(t))^2 dt$ doesn't make the inspiratory resistance power smaller than (25). The above assumption is therefore wrong. Consequently, the inspiratory flow constant function (19) together with the associated resistance power (25) proves as the minimal solution. Interestingly, none of the authors, who have chosen the respiratory rate as their research subject, used this result.

From Eq. (15) it also follows that (19) (at least in humans) is about 5 times smaller than $\left|\dfrac{dV}{dt}\right|$ at the beginning of expiration, i.e. at time $t = 0$ in (13) which is why (19) cannot have a turbulent character and therefore Eq. (9) and Eq. (18) could also be used for inspiration.

The linear inspiratory volume function $V(t) = FRC + \dfrac{V_T}{T_I} t$ as shown in figure 4 then follows from Eq. (19) by integration with the initial condition $V(0) = FRC$ which corresponds to the intrapulmonary volume at the end of the previous expiration.

The postulate (P) therefore leads, with spontaneous breathing at rest, to the constant inspiratory volume flow (19) which is accordingly to be assumed here in the following. However, a discussion about this now power-minimized inspiratory volume flow pattern could still develop in subsequent publications and (19) will actually be modified slightly below.

## *Minimum of the entire physical respiratory power*

Because of the passivity of the spontaneous expiration, the organism no longer has to provide (18) in this breathing phase. It follows from this that for the expansion from (25) to the entire breathing period $[0, T_I + T_E]$, only the addition of zero expended resistance power during the passive expiratory duration $T_E$ is necessary. This is taken into account in the following by multiplying Eq. (25) with $\dfrac{T_I}{T_I + T_E} = F \cdot T_I$ which corresponds to an averaging of the power (25) over the entire breathing period.

The entire physical respiratory power $P$ to be expended by the organism is thus composed of the two components (6) and the averaging of (25): $P = \dfrac{V_T^2}{2C} \cdot F + R \cdot \left(\dfrac{V_T}{T_I}\right)^2 \cdot F \cdot T_I$ and with Eq. (20) it follows:

$$P = V_T^2 \cdot \left( \dfrac{F}{2C} + F^2 \cdot R \cdot \dfrac{I:E+1}{I:E} \right). \qquad (26)$$

The variable $V_T$ can be substituted for the respiratory rate $F$ with Eq. (22) (or vice versa). The first choice seems to be simpler. It then follows for the entire respiratory physical power to be provided by the organism:

$$P = \left( V_D + \dfrac{\dot{V}CO_2}{F \cdot F_{Et}CO_2} \right)^2 \cdot \left( \dfrac{F}{2C} + F^2 \cdot R \cdot \dfrac{I:E+1}{I:E} \right) \qquad (27)$$

This equation was not published until now. From a physical point of view the side condition $P \geq 0$ is obvious. At this point one could ask whether there is an optimal compliance $C$ (modulo the discussion in the second subsection) and an optimal resistance $R$. The answer is quite simple: $C = \infty$ and $R = 0$, since both together lead to $P = 0$. This result of course is biologically impossible. By the way, $C = \infty$ in Eq. (4) gives $W_C = 0$, hence a passive expiration would not be possible any more. Animals with a body covered by fur have limited ability to sweat, relying heavily on panting. A $V_T$ close to $V_D$ and high $F$ is characteristic of panting causing more air to ventilate the dead-space $V_D$ in order to increase heat loss by water evaporation without increasing $CO_2$-loss. Due to this physiological dead space ventilation and of course due to the huge necessary gas exchange surface as well which requires a bronchial tree, the same impossibility applies to the optimal dead space $V_D = 0$ and thereof consequently the mentioned impossibility of $R = 0$. Questions like these seemingly are irrelevant in this context. Therefore, the parameters $R$, $C$ and $V_D$ are given by the anatomy (or pathophysiology) and consequently constant within this context. The $CO_2$-homeostasis of the organism requires an (at least largely) constant $F_{Et}CO_2$ and $\dot{V}CO_2$ is also constant (with the already mentioned restrictions) under spontaneous breathing at rest. So far, the respiratory rate $F$ and the $I:E$-ratio have been considered constant as well. This should initially continue to apply to the $I:E$-ratio. After this little sense of achievement, the question arises as to which the breathing rate $F$ in the sense of (P) could be optimal. In order to obtain the minimum of the respiratory power, (27) must be derived by the respiratory rate $F$ in order to then calculate the optimal respiratory rate $F_{opt}$ from $\frac{dP}{dF} = 0$. It is then to be checked whether this leads to a minimum of Eq. (27) and, hopefully, not to a maximum. Let $\eta$ be the constant energy conversion efficiency of respiration (Campbell 1957) at rest, then from $\frac{dP}{dF} = 0$ of course $dP_{resp}/dF = 0$ follows as well, since $P_{resp} = P/\eta$. Consequently (as discussed above) $\dot{V}CO_2$ is largely independent of $F$ within a neighbourhood of $F_{opt}$ as well. Assuming that $\eta \geq 0.22$, for example the human data in table 2 and the following result then show that $P_{resp}$ accounts for less then 0.25% of the basal metabolic rate in this simple model of spontaneous breathing at rest.

To keep it general, all the 6 parameters $V_D$, $I:E$-ratio, $\dot{V}CO_2$, $F_{Et}CO_2$, $C$ and $R$ in Eq. (27) have not yet been specified, since these parameters are not human-specific. (The $I:E$-ratio will be discussed below.) The results obtained should therefore apply more widely in biology, possibly even for most lung-breathing animals with a thorax, especially mammals. A whale or other mammals who predominantly live in water also use the buoyancy of the air in the lungs which admittedly still would have to be included in the above considerations.

After deriving, multiplying by $F^2$ and then setting equal to zero, Eq. (27) results in a 3rd degree polynomial in the breathing rate $F$, whose positive real zero $F_{opt}$ has to be calculated:

$$0 = F^3 \cdot 2V_D^2 \cdot R \cdot \frac{I:E+1}{I:E} + F^2 \left( \frac{V_D^2}{2C} + \frac{2V_D \cdot \dot{V}CO_2 \cdot R}{F_{Et}CO_2} \frac{I:E+1}{I:E} \right) - \frac{\dot{V}CO_2^2}{2F_{Et}CO_2^2 \cdot C} \quad (28)$$

Every polynomial of *n*-th degree is known to have *n* solutions, so the above equation has 3 solutions which could also be complex. Undoubtedly, only a positive real solution is physiologically meaningful. The polynomial (in the variable *F*) on the right-hand side of Eq. (28) has only real coefficients and as a real polynomial of degree 3 it has at least one real solution.

The question that immediately arises: Is there a positive solution at all? This question can be answered with yes, because: The multiplication with $F^2$ could have brought the trivial solution $F = 0$. This multiplication was allowed, however, since $F = 0$ is biologically only possible for $\dot{V}CO_2 = 0$ and due to $RQ \neq 0$ ultimately only for $\dot{V}O_2 = 0$. The latter is indeed a solution for living beings for a short period of time that hibernate (e.g. hamster) and take long pauses in breathing in this state. (Side note: Since not only the metabolism but also the urine excretion is reduced in hibernation, the above approximation of the $CO_2$ balance also applies in hibernation.) However, a human with $F = 0$ is subject to resuscitation, but lowering the body temperature also reduces the oxygen consumption $\dot{V}O_2$ in humans substantially, so that $F = 0$ is possible for a short time during hypothermic cardiac arrest in the cardiac operating room, but not under spontaneous breathing as in hibernators.

If $F$ in (28) increases starting from 0, then the first two summands always remain positive increasing and together reach the value $\dfrac{\dot{V}CO_2^2}{2 F_{Et}CO_2^2 \cdot C}$ exactly once which is why Eq. (28) has exactly one positive real solution provided that $\dot{V}CO_2 > 0$. This solution is therefore unique which is physiologically reasonable, because: Assuming Eq. (28) actually had 3 positive real solutions, then the organism would be spoiled for choice between 3 different optimized breathing rates and even worse, evolution would not have focused on one stable respiratory rate, but could have established a metastable level.

Every polynomial of degree less than or equal to 4 can be solved with radicals. A simple solution of Eq. (28) is $F = \dfrac{-\dot{V}CO_2}{V_D \cdot F_{Et}CO_2}$ which cannot be the biological one. However, dividing of Eq. (28) by $F + \dfrac{\dot{V}CO_2}{V_D \cdot F_{Et}CO_2}$ results in a quadratic polynomial and the unique real positive root thereof is:

$$F_{opt} = \frac{1}{8}\left( \sqrt{\frac{I:E}{I:E+1} \cdot \frac{1}{R \cdot C} \cdot \frac{16 \cdot \dot{V}CO_2}{V_D \cdot F_{Et}CO_2} + \left(\frac{I:E}{I:E+1} \cdot \frac{1}{R \cdot C}\right)^2} - \frac{I:E}{I:E+1} \cdot \frac{1}{R \cdot C} \right) \quad (29)$$

This equation has never been published before as well. The second derivative of (27): $\dfrac{d^2 P}{dF^2} = F^{-3} \cdot \dfrac{\dot{V}CO_2^2}{C \cdot F_{Et}CO_2^2} + 2 \cdot V_D^2 \cdot R \cdot \dfrac{I:E+1}{I:E}$ has no negative coefficients and since the biological relevant solution (29) is always positive (or in the extreme case almost 0), this $F_{opt}$ happily is a minimum of the power (27). The associated tidal volume $V_{T,opt}$ is then to be calculated from Eq. (22).

Since $\dot{V}CO_2$ remains constant under spontaneous breathing at rest with the respiratory rate $F_{opt}$, the minute volume $F_{opt} \cdot V_{T,opt}$ is also constant under the requirement of a physiological $CO_2$ homeostasis and thus a largely constant $F_{Et}CO_2$. With Eq. (19) such ventilation is called constant flow volume-controlled which suggests the function $V(t)$ in figure 6 as an idealized simple model of spontaneous breathing at rest. Surely there exists no simpler one with comparable results.

## *Acceleration power and other neglects*

A secondary result during the derivation of the Hagen-Poiseuille law is the fact that the maximum velocity in the middle of a laminar flow in a pipe is exactly twice as large as the mean flow velocity $\bar{v}$. It is known that the velocity of the breathing gas is highest in the main bronchi and the trachea. Let $r$ be the radius of the trachea (Pinkerton 2015) (in human adults $\approx 13$ mm), then because of (19) $\bar{v} = \dfrac{V_{T,opt}}{T_I \cdot r^2 \pi}$ is the mean velocity in the trachea during spontaneous inspiration at rest. In order to accelerate the breathing gas with the mass *m* to the maximum speed $v = 2\bar{v}$ the energy $\dfrac{m \cdot v^2}{2}$ is required. The inspiratory acceleration power necessary to reach the kinetic energy $\frac{(2 \cdot \overline{v})^2 \cdot \rho \cdot V_T}{2}$ of the breathing gas with the density $\rho$ for every breathing period, even at the maximum velocity $2\overline{v}$ for air ($\rho \approx 1.2 \, kg/m^3$), is (as a simple calculation shows using the human data in table 2) at least a factor of $10^5$ less than (27) and could therefore be neglected.

The acceleration of the lungs and chest wall between the breathing phases is comparably low, because the accelerated mass is (compared to the breathing gas) indeed a factor of $10^4$ larger, but the square of the maximum velocity is about the same factor $10^4$ less.

As mentioned before, nose, pharynx, throat, trachea and bronchi are surrounded by bones or cartilage (Peake 2015) which are mainly responsible for $R$, while the bronchioles only make a very small contribution to this (West 2012). Therefore, analogous to the lungs themselves, this dead space at the transition between the breathing phases holds back a very small volume $\Delta V$ and accordingly influences the volume flow. The expiratory pressure gradient (14) is almost zero at the end of expiration, however due to Eq. (9) the volume flow (19) causes an inspiratory resistance pressure gradient $p_R = R \cdot \frac{V_T}{T_I} = \frac{\tau}{T_I} \cdot p_C$ which, due to Eq. (5) and Eq. (15), is about 5 times smaller than the compliance pressure (5). Nevertheless, this pressure gradient $p_R$ is still found at the end of inspiration mainly in this inelastic (i.e. rather stiff) part of the dead space with an admittedly very low (here negligible) compliance $C_D \approx 0$, at least $C_D << C = C_{Lungs\&Thorax}$. With a constant body temperature, an isothermic process can be assumed within the lungs, hence the Boyle-Mariotte law is applicable within the dead space, consequently $(V_D + \Delta V) \cdot p_B = V_D \cdot (p_B + p_R)$ approximately holds, whereof $\Delta V = \frac{V_D \cdot p_R}{p_B} = \frac{V_D \cdot R \cdot V_T \cdot F_{opt}}{p_B} \left( \frac{I:E+1}{I:E} \right)$ follows. For example the adult human data in table 2 give $\Delta V = 0.14 \, ml$. Hence, in retrospect Kirchhoff's current (which corresponds in respiratory physiology to Kirchhoff's volume flow) law was at least approximately applicable. Even if $C_D$ is not negligible and $\Delta V$ would be 5 to 10 times larger, the inspiratory resistance pressure gradient can not significantly affect (22) and (29). Nevertheless, the inspiratory volume flow is no longer constant at the very beginning and the very end of inspiration, rather the corners and edges in figure 6 like a "mollifier" are rounded off, so that the volume curve comes closer to that shown in physiology textbooks (Hall 2020).

With all the neglects and approximations made so far, no exact result can be expected by Eq. (29). The art of applied mathematics in the natural sciences is to choose such neglects and approximations that the resulting as simple as possible formalism still describes nature as well as possible. In subsequent publications, however, these approximations could be analyzed more precisely using more complex models.

## Example: The respiratory rate of an adult human and a human newborn baby

In the literature, $\dot{V}CO_2 = \dot{V}O_2 \cdot RQ$ is rarely given, which is why in table 2 this parameter was calculated from the $O_2$-consumption rate $\dot{V}O_2$ with the respiratory quotient $RQ$. Using the example of humans, the results in the penultimate and last column in table 2 should reinforce confidence in the concept (P) that led to (29). Moreover, for the newborn human there are (apart from descriptive studies) no clinical studies at all which predict or estimate the respiratory rate at rest. However, the present publication only claims to be a theoretical concept for the spontaneous breathing rate in the resting state. The compliance $C$ increases with body weight and age (Bolle 2008; Gaultier 1984; Rendas 1978) as can be seen in table 2 as well. However, in the absence of (most of the other comparable) animal data analogous to table 2, this publication is intended to encourage Eq. (29) further scientific tests.

|  | $V_D$ ml | $F_{Et}CO_2$ % | $\dot{V}CO_2$ ml/min | $I:E$ ratio | $R$ mbar/(l/s) | $C$ ml/mbar | $F_{opt}$ 1/min | $V_{T,opt}$ ml |
|---|---|---|---|---|---|---|---|---|
| Adult human | 160 | 5.2 | 196.8 | 0.5 | 3 | 100 | 13.2 | 446.7 |
| Newborn human | 7.2 | 5.2 | 19.1 | 0.67 | 30 | 3 | 33.9 | 18.05 |

**Table 2:** The data dead space $V_D$, end-tidal CO$_2$-fraction $F_{Et}CO_2$, CO$_2$ production rate of the organism $\dot{V}CO_2$, duration of inspiration : duration of expiration $I:E$-ratio, airway resistance $R$, static compliance $C$ of the lungs and thorax close to the $FRC$ within this table are derived from the literature (Galetke 2007; Long 2017; Butler 1957; Galetke 2007; Guo 2005; Huang 2016; Levitzky 2018; Pasquis 1976; Boggs 1984; Weingarten 1990; Lumb 2016b; Joehr 1993). The respiratory rate $F_{opt}$ of an adult human and a newborn baby weighing about 3.5 kg in the penultimate column is the solution of equation (29) using these data. For example the adult $F_{opt}$ gives

$$\frac{1}{8}\left( \sqrt{\frac{0.5}{0.5+1} \cdot \frac{1000 \cdot 60}{3 \cdot 100} \cdot \frac{16 \cdot 196.8}{160 \cdot 0.052} + \left( \frac{0.5}{0.5+1} \cdot \frac{1000 \cdot 60}{3 \cdot 100} \right)^2} - \frac{0.5}{0.5+1} \cdot \frac{1000 \cdot 60}{3 \cdot 100} \right) \approx 13.2 / \min.$$

The Factor $1000 \cdot 60$ is the conversion coefficient from ml. to liter respectively sec. to min. The associated tidal volume $V_{T,opt}$ in the last column follows from (22) and the product $F_{opt} \cdot V_{T,opt}$ is then the related minute volume. The above data allow the calculation of the entire physical respiratory power (27) of the adult human (30.6 mW) and the human newborn (3.8 mW) as well.

## Quantitative influencing factors on the respiratory rate

At least as interesting, is the following question: How big is the influence of the 6 parameters $X \in \{V_D, F_{Et}CO_2, \dot{V}CO_2, I:E, R, C\}$ in Eq. (29) on the optimal $F_{opt}$, or to put it somewhat more mathematically while keeping it dimensionless: How does $\frac{\Delta F_{opt}}{F_{opt}}$ change, if only one of the 6 parameters $\frac{\Delta X}{X}$ changes slightly which leads to the linear ansatz: $\frac{\Delta F_{opt}}{F_{opt}} = k_X \cdot \frac{\Delta X}{X}$, in which $F_{opt}$ and the parameters $X$ which led to the solution $F_{opt}$, but not $\Delta F_{opt}$ and $\Delta X$, are regarded as constant. It is now necessary to calculate these 6 dimensionless factors $k_X$. A little rewritten $\frac{\Delta F_{opt}}{\Delta X} = k_X \cdot \frac{F_{opt}}{X}$ and for $\Delta X \to 0$ follows $\frac{\partial F_{opt}}{\partial X} = k_X \cdot \frac{F_{opt}}{X}$ where $\frac{\partial F_{opt}}{\partial X}$ is the partial derivative of $F_{opt}$ according to the parameter $X$ at the point $X$ which led to $F_{opt}$.

For example: $\frac{\partial F_{opt}}{\partial (\dot{V}CO_2)} = \frac{I:E}{I:E+1} \cdot \frac{1}{R \cdot C} \cdot \frac{1}{V_D \cdot F_{Et}CO_2} / \sqrt{\frac{I:E}{I:E+1} \cdot \frac{1}{R \cdot C} \cdot \frac{16 \cdot \dot{V}CO_2}{V_D \cdot F_{Et}CO_2} + \left(\frac{I:E}{I:E+1} \cdot \frac{1}{R \cdot C}\right)^2}$

Note: Since $\dot{V}CO_2$ is largely constant within a small neighbourhood of $F_{opt}$ the above partial derivative is surely valid for the parameter $k_{\dot{V}CO_2} = \frac{\dot{V}CO_2}{F_{opt}} \cdot \frac{\partial F_{opt}}{\partial(\dot{V}CO_2)}$.

**Claim:** Since the parameters $C$ and $R$ occur in Eq. (29) exclusively together in form of the product $C \cdot R$, the following holds: $\frac{C}{F_{opt}} \cdot \frac{\partial F_{opt}}{\partial C} = \frac{R}{F_{opt}} \cdot \frac{\partial F_{opt}}{\partial R}$. The same symmetry is true for $\dot{V}CO_2$, $1/F_{Et}CO_2$ and $1/V_D$ as well and consequently $\frac{\dot{V}CO_2}{F_{opt}} \cdot \frac{\partial F_{opt}}{\partial(\dot{V}CO_2)} = -\frac{V_D}{F_{opt}} \cdot \frac{\partial F_{opt}}{\partial V_D} = -\frac{F_{Et}CO_2}{F_{opt}} \cdot \frac{\partial F_{opt}}{\partial(F_{Et}CO_2)}$ holds too.

Proof: Let $x = A \cdot B$, then $\frac{\partial F(x)}{\partial A} = \frac{\partial F(A \cdot B)}{\partial A} = \frac{dF}{dx} \frac{\partial x}{\partial A} = \frac{dF}{dx} B$ consequently $\frac{dF}{dx} = \frac{1}{B} \frac{\partial F(x)}{\partial A}$ and analogously $\frac{dF}{dx} = \frac{1}{A} \frac{\partial F(x)}{\partial B}$. Therefore, $A \frac{\partial F(x)}{\partial A} = B \frac{\partial F(x)}{\partial B}$ and dividing by $F_{opt}$ gives the first result. Let now $x = \frac{A}{B}$, then only a slightly changed proof gives $A \frac{\partial F(x)}{\partial A} = -B \frac{\partial F(x)}{\partial B}$ and again dividing by $F_{opt}$ gives the negative sign result.

**Corollary**: $k_R = k_C$ and $k_{\dot{V}CO_2} = -k_{V_D} = -k_{F_{Et}CO_2}$ (s. table 3).

Hence, strictly speaking, equation (29) is (only) a function of the 3 parameters $\tau = R \cdot C$, $I:E$ and $\frac{\dot{V}CO_2}{V_D \cdot F_{Et}CO_2}$, a fact that can of course be seen directly in Eq. (29) as well.

**Claim:** Let $\dot{V}CO_2 > 0$, $k = \frac{16 \cdot \dot{V}CO_2}{V_D \cdot F_{Et}CO_2} > 0$, $R \cdot C > 0$ and $(I:E) > 0$, then $\frac{I:E}{F_{opt}} \cdot \frac{\partial F_{opt}}{\partial(I:E)} > 0$.

Proof: Let $f(x) = \frac{x}{x+1}$, then $f(x)$ is a strictly increasing (at least) $C^1$ function for $x > 0$ and moreover $f(x) > 0$. Let now $g(x) = \frac{1}{8}\left(\sqrt{x \cdot k + x^2} - x\right)$, then $g(x)$ has the same properties. Due to the chain rule, the composition of $C^1$ functions is again a $C^1$ function. Moreover, the composition of strictly increasing functions is again a strictly increasing function, consequently $g(f(x))$ is a strictly increasing (at least) $C^1$ function for $x > 0$. Hence, because $F_{opt} > 0$ (29) as a function of the parameter $I:E$ has the same property.

**Corollary:** Within the biologically relevant range $(I:E) > 0$ the function (29) $F_{opt} = F_{opt}(I:E)$ is strictly positive if $\dot{V}CO_2 > 0$. Consequently the $I:E$-ratio "correlates" positively with $F_{opt}$ (s. table 3).

Similar proofs can be used for the other parameters $\frac{X}{F_{opt}} \frac{\partial F_{opt}}{\partial X}$ at least within a small neighbourhood close to all the biologically relevant $X$ (for humans given in table 2).

| $X$ | $V_D$ | $F_{Et}CO_2$ | $\dot{V}CO_2$ | $I:E$ | $R$ | $C$ |
|---|---|---|---|---|---|---|
| Adult human | -0.69 | -0.69 | 0.69 | 0.21 | -0.31 | -0.31 |
| Newborn human | -0.75 | -0.75 | 0.75 | 0.15 | -0.25 | -0.25 |

**Table 3** Dimensionless factors $k_X = \dfrac{X}{F_{opt}} \cdot \dfrac{\partial F_{opt}}{\partial X}$ at the point $X$ of these parameters given in table 2. Because of all $\dfrac{X}{F_{opt}}$ are positive, the sign of $k_X$ is a parameter for the "correlation" direction between $F_{opt}$ and $X$. The parameters $\dot{V}CO_2$ and $I:E$–ratio therefore "correlate" positively with $F_{opt}$, with $\dot{V}CO_2$ having a significantly stronger effect on $F_{opt}$. All other parameters "correlate" negatively with $F_{opt}$. Due to symmetries of Eq. (29) $k_{\dot{V}CO_2} = -k_{V_D} = -k_{F_{Et}CO_2}$ and $k_R = k_C$ applies.

Concerning the human data in table 2, the quantitative effect in terms of the $k_X$ of these 6 parameters $X$ on $F_{opt}$ is summarized in table 3 which is the basis for the first sentence in the abstract. If one only looks at the sign of the $k_X$, then (because all $\dfrac{X}{F_{opt}}$ are positive) an increase in $\dot{V}CO_2$ or $I:E$-ratio increases the respiratory rate $F_{opt}$ with $\dot{V}CO_2$ having a significantly stronger effect on $F_{opt}$, whereas the parameter $V_D$, $F_{Et}CO_2$, $R$ and $C$ "correlate" negatively with the respiratory rate $F_{opt}$. This explains the relatively high quiet breathing rate $F_{opt}$ of a human newborn compared to an adult: A newborn produces about twice as much $CO_2$ per kg body weight and time unit due to the approximately twice as high $\dot{V}O_2$ (Long 2017), whereby the compliance $C$ is considerably lower (Huang 2016; Kerem 1996). The relatively large head is responsible for a somewhat larger $\dfrac{V_D}{V_{T,opt}}$-ratio (Numa 1996), hence the relatively larger anatomical dead space $V_D$ as well as the relatively higher resistance $R$ reduce the breathing rate $F_{opt}$.

Again using the example of adult humans then Eq. (15) together with Eq. (21) gives a dimensionless factor $\tau \cdot F \cdot (I:E+1) = q$ and the data from table 2 show that:
$$\tau_{Newborn} \cdot F_{opt,Newborn} \cdot (I:E_{Newborn}+1) < \tau_{Adult} \cdot F_{opt,Adult} \cdot (I:E_{Adult}+1) \qquad (30)$$
In adults it is noted that the $I:E$-ratios tend to be approximately the same $\approx 1:2$ inter-specifically (Boggs 1984). The time constants $\tau$ are of course given by the anatomy, physiology and the age. The parameters $F_{opt,Newborn}$ and / or the $I:E_{Newborn}$-ratio in (30) could therefore accordingly be increased without adversely affecting expiration (12) since $q = \dfrac{\tau}{T_E}$. As shown above, $F_{opt}$ is a continuous strictly increasing function of the parameter $I:E$ within the biologically relevant range $(I:E) > 0$ provided that $\dot{V}CO_2 > 0$. Consequently there must be exactly one $I:E$-ratio for the (non hibernating) newborn, so that the equal sign applies in Eq. (30). However, this solution can only be calculated numerically. With the daring hypothesis that this equal sign applies in Eq. (30) and the right side of Eq. (30) defines a dimensionless number $q = \tau \cdot F \cdot (I:E+1)$ for each species, the $I:E$-ratio and $F_{opt}$ for the newborn or the child could be calculated from Eq. (30). For the human newborn the $F_{opt,Newborn}$ would then slightly increase to 35/min which is still within the normal range. At the same time the respiratory power (27) of the newborn would then even decrease by about 4.6%. Side-note: Concerning the $I:E$-ratio none of the authors, who have chosen the respiratory rate as their subject, used this parameter.

## Discussion in the light of previous publications

In 1950, ventilation was only possible with the historical iron lung, e.g. the Drinker respirator, in which test subjects were ventilated as part of a study (Otis 1950). The authors used clinical data, in some cases even with added $CO_2$, to minimize the respiratory power with mathematical aids in order to calculate an optimal respiratory rate as well, assuming that the breathing pattern is a sine wave. However, their method cannot be generalized in a comparable way which is why the respiratory rate of a newborn human could not be calculated, especially since no newborns were included into the study. In addition, the power unit $G \cdot M \cdot CM/MIN$ selected in this study is hardly comprehensible today and, according to today's standards, would no longer have the physical unit of a power (Schramm 2010). In this publication (and also subsequent publications) there is also a term like the one in (17) which, as shown above, no longer allows the concept of a time constant $\tau$. Moreover, this publication does not reveal to what extent the discussed parameters ultimately influence the resting respiratory rate. The term lung respective thorax compliance was not yet developed in its present form in 1950, however, the elastic work was already known by nearly the same authors (Rahn 1946). The mathematical methods, such as the Euler-Lagrange formalism, were fully developed at this time, but the calculus of variations was not used.

In a similar clinical study based on (Otis 1950), Mead (Mead 1960) examined both guinea pigs and human subjects, but ultimately came to the conclusion that the optimal respiratory rate is not achieved by minimizing respiratory power, but rather by the force to be applied to the respiratory muscles. Crosfill and Widdicombe (Crosfill 1961) have each derived an equation for minimizing respiratory performance and an equation for minimizing the mean force to be exerted by the respiratory muscles by assuming the alveolar ventilation $F \cdot (V_T - V_D)$ to be constant. However $F \cdot (V_T - V_D)$ should have been included in the derivation of the optimal breathing rate as a side condition. Bates and Milic-Emili (Bates 1993) set out to determine if the viscoelastic properties of the lung could better refine the determinants of optimal breathing frequency, however, during quiet breathing as well as exercise. They simulated the lung using computational methods. To their surprise, they concluded that the viscoelastic properties of the respiratory system tissues do not significantly alter the prediction of optimal breathing frequency in humans or dogs and inspiratory work was a better predictor of optimal breathing frequency than an inspiratory pressure-time integral.

The quantitative measurement of $CO_2$ was already described by Tyndall in 1862 (Tyndall 1862) and the infrared gas analysis of air (Luft 1943) in 1943, but at that time it was still not used in biology or medicine for the continuous measurement of the $F_{Et}CO_2$ (Jaffe 2008) or, if at all, only with great technical effort (Miller 1950) which is why the above publications have not included (1) in their considerations. Noël (Noël 2019) has integrated the gas transport into his model, however, thus buying in other difficulties and was therefore unable to answer all the questions asked. Eitan Kerem (Kerem 1996) completely dispensed with mathematics in order to describe qualitative influences of the parameters $V_D$, $\dot{V}CO_2$, $R$, $C$, $paCO_2$ but also other possible influencing factors on the respiratory rate in small children and newborns, so quantitative data are completely lacking. With regard to the volume flow Hancao (Hancao 2012) has chosen a minimizing approach for both the inspiratory and the expiratory airflow in a non-linear multi-compartment numerical model. Under additional assumptions, he achieved results for the respiratory flow that differ completely from (13) and (19).

## Limits under physical exercise

As already mentioned, expiration is passive under quiet spontaneous breathing. This assumption generally no longer applies under physical exercise, because in the latter case the respiratory gas (usually air) is actively exhaled via the respiratory muscles. In order to maintain a physiological $CO_2$ homeostasis in the organism due to (1), an increased alveolar ventilation $F \cdot (V_T - V_D)$ is necessary with increased physical performance because of the excess $CO_2$ production rate $\dot{V}CO_2 = \dot{V}O_2 \cdot RQ$ coupled to the aerobic metabolism. As everyone knows, this requires an increase in the tidal volume $V_T$ (via the inspiratory and expiratory reserve volume) as well as the respiratory rate $F$. Because of Eq. (21) the higher respiratory rate $F$ shortens the expiration time $T_E$ and due to Eq. (12) both, the increased tidal volume $V_T$ and the shorter $T_E$ are responsible for the active accelerated expiration by increasing the intrathoracic pressure $p_P$ by the respiratory muscles. Therefore, expiration is no longer passive under an appropriate higher level of physical performance. Incidentally, the expiratory reserve volume can never be exhaled without the aid of the respiratory muscles. For this reason (28) can only apply in an (at least approximately) resting state.

Another limit of Eq. (28) is the following: It is known that the lactate concentration within the organism increases under heavy physical exertion which leads to a metabolic acidosis shifting of the balance $CO_2 + H_2O \rightleftarrows H_2CO_3 \rightleftarrows H^+ + HCO_3^-$ to the left and thus causing an increased $CO_2$ release from the bicarbonate buffer which then has to be included in the $CO_2$ balance addressed in the beginning. This is why, even the $RQ$ can sometimes rise above the value 1 independent of nourishment, however only when almost physical exhaustion occurs.

## Possible applications in medicine

Since all schoolchildren already know their breathing rate at rest, the practical benefit of the resting breathing rate is low and the present publication only appears to be of academic interest. The parameters $F_{opt}$, $F_{Et}CO_2$, $I:E$-ratio which occur in (29) can be measured with simple aids and also $\dot{V}CO_2$, $C$ and $R$ (the latter with a body plethysmograph) can still be recorded without much effort. Moreover, by using Eq. (1), one of the parameters $F_{Et}CO_2$, $\dot{V}CO_2$ or $V_D$ can be substituted in Eq. (29) by the easily measurable tidal volume $V_T$ and the dead space $V_D$ can be determined from the Enghoff-modified Bohr equation (Enghoff 1938; Bohr 1891). If 6 of the 7 parameters (including $F_{opt}$) in Eq. (29) are known, then the 7th can be calculated therefrom which is why equation (29) can be of interest not only in physiology and biology, but under certain circumstances also in medicine. Since $C$ and $R$ are routinely measured in a body plethysmograph, Eq. (29) could, for example, extend this pulmonary instrument.

It cannot be ruled out that this equation could also become valid in parts of the pathophysiology which, however, would have to be proven in clinical studies: It is known that the airway resistance $R$ is increased during an asthma attack which results in an extended time constant $\tau$ and leads due to Eq. (12) to an extended expiratory duration. The time $T_E$ is then no longer sufficient to passively exhale the tidal volume $V_T$ consequently $FRC \neq V(p_B)$ if the $FRC$ is still defined to be the volume in the lungs at the end of expiration and $V(p)$ is still the static function, as introduced in the beginning. In this case, under ventilation, the increased intrapulmonary volume $FRC - V(p_B)$ is called trapped volume. Because of Eq. (8) this leads to a $PEEP$ which is then referred to as intrinsic $PEEP$ (sometimes called auto-$PEEP$). In order to counteract this, the patient has no choice but to actively expire during an asthma attack, i.e. supported by the respiratory muscles (and the $PEEP$) which is why Eq. (28) can no longer be valid in an asthma attack. However, an asthma attack in the body plethysmograph is the exception anyway. A comparable pathophysiology can be observed in COPD patients. In COPD patients, however, the permanently high airway resistance $R$ and moreover, a permanently increased compliance $C$ (due to a loss of elasticity of the lungs) leads to a permanently extended time constant $R \cdot C$. The resulting (due to Eq. (13)) permanently lower expiration flow $\frac{dV}{dt}$ leads to a permanent increased $FRC$ with all its consequences such as high intrathoracic pressure, barrel chest, ... therefore equation (28) would have to be modified accordingly for COPD patients. Moreover, in a COPD diseased lung, certain portions may be more affected than others, therefore $\tau$ is not homogeneous within the lungs any more. Consequently, the alveolar pressure cannot be a common pressure for all the alveoli any more and pressure equalization during and between the breathing phases is the consequence, requiring additional breathing work.

It can be assumed that hypoxia from a pulmonary origin affects the minute volume $F \cdot V_T$ in a similar way as described by Duffin in altitude medicine (Duffin 2007). For patients with lung diseases without intrinsic $PEEP$ and still sufficiently high arterial O$_2$-partial pressure, however, nothing should diminish the applicability of Eq. (29). Pneumonia, for example, reduces the compliance $C$ (Somerson 1971) and the accompanying rise in body temperature increases $\dot{V}O_2$ and thus $\dot{V}CO_2$. Both parameters lead to an increase of $F_{opt}$, a clinically age-old finding.

Since the organism prefers a normal $pH$ value instead of a CO$_2$-homeostasis which in pathophysiology is referred to as the respiratory compensation of a metabolic disorder, there are also $F_{Et}CO_2$ deviations from the norm value which due to Eq. (1) and Eq. (29) accordingly affect $F \cdot V_T$ without primary lung diseases (Kußmaul 1874). Hence, metabolic disorders might be applications of Eq. (29) as well.

The applicability of Eq. (29) for these pathophysiologies of course should be first evaluated in clinical studies.

As already mentioned in the theoretical subsection, the "hypothetic time constant function" $\tau(V)$ could partially explain the course of spirometry functions in medicine.

# Final remarks
Although the translation of respiratory physiology into a primarily physical and mathematical language together with a repeatedly applied postulate from the theory of evolution using the example of humans has led to expected results, it must not be overlooked that evolution has spent millions of years on a human being and the surrounding wildlife. Even if the breathing mechanics is only a very tiny part of it, the attempt to describe this part in a few lines of evolution, biology, physiology, physics and mathematics can just remain an attempt, everything else would be overconfident. But if this publication helps to better understand the mechanics of breathing, especially spontaneous breathing at rest, or perhaps even to stimulate further reflection and publication, it will perhaps help that not only the physiology of humans, but also the biology of many living beings to become a tiny bit richer.

# Abbreviations

| | |
|---|---|
| $C(p)$ | Static compliance = $dV/dp$ |
| $C$ | Static compliance at the inflection point of $V(p)$ |
| $F$ | Respiratory rate = $1/(T_I + T_E)$ |
| $F_{opt}$ | Power-minimized respiratory rate at rest |
| $FCO_2$ | $CO_2$-fraction = $pCO_2/p_B$ |
| $F_iCO_2$ | Inspiratory $CO_2$-fraction |
| $F_{Et}CO_2$ | Endtidal $CO_2$-fraction |
| $FRC$ | Functional residual capacity |
| $I:E$ | Inspiratory duration / expiratory duration = $T_I/T_E$ |
| $m$ | Mass |
| $p$ | Pressure |
| $p_C$ | Compliance pressure |
| $p_R$ | Resistance pressure |
| $p_B$ | Ambient pressure |
| $P$ | Physical power |
| $PEEP$ | Postitive end-expiratory pressure |
| $P_C$ | Compliance power |
| $P_R$ | Resistance power |
| $R$ | Airway-resistance |
| $RQ$ | Respiratory quotient = $\dot{V}CO_2/\dot{V}O_2$ |
| $RV$ | Residual volume |
| $t$ | Time |
| $\tau$ | Time constant = $R \cdot C$ |
| $T_I$ | Inspiratory duration |
| $T_E$ | Expiratory duration |
| $TC$ | Total lung capacity |
| $v$ | Velocity |
| $\bar{v}$ | Mean velocity |
| $V$ | Volume |
| $V(p)$ | Static "pressure-volume" function |
| $\dot{V}O_2$ | $O_2$-consumption rate |
| $\dot{V}CO_2$ | $CO_2$-production rate = $RQ \cdot \dot{V}O_2$ |
| $V_T$ | Tidal volume |
| $V_{T,opt}$ | Power-minimized tidal volume at rest |
| $V_D$ | Dead space |
| $dV/dt$ | Volume flow |
| $W_C$ | Compliance work |
| $W$ | Physical work |

Abbreviations